%% file: tmlr.tex
\documentclass[10pt]{article} % For LaTeX2e
\usepackage[preprint]{tmlr}

\input{math_commands.tex}

\usepackage{hyperref}
\usepackage{url}
\usepackage{graphicx}
\usepackage{booktabs}
\usepackage{multirow}
\usepackage{float}
\usepackage{wrapfig}
\usepackage{xspace}
\usepackage{caption}
\usepackage{subcaption}
\usepackage[edges]{forest}
\usepackage[table]{xcolor}
\usepackage[most]{tcolorbox}
\usepackage{enumitem}

\input{boxs}

\definecolor{tabgreen}{RGB}{141, 211, 199}
\definecolor{tabred}{RGB}{251, 128, 114}

\usepackage{siunitx}
\newcommand{\ie}{i.e.,\xspace}
\def\figref#1{Figure~\ref{fig:#1}}

\def\figsref#1#2{Figures~\ref{fig:#1}-\ref{fig:#2}}

\def\tabref#1{Table~\ref{tab:#1}}

\def\eqref#1{Eq.~\ref{eqn:#1}}

\def\Appref#1{Appendix~\ref{sec:#1}}
\def\seclabel#1{\label{sec:#1}\label{p:#1}}
\def\secref#1{\S\ref{sec:#1}}

\title{Insights from the ICLR Peer Review and Rebuttal Process}
\author{\name Amir Hossein Kargaran$^{1,3}$\thanks{Equal contribution.} \quad
  Nafiseh Nikeghbal$^{2,3}$\footnotemark[1] \quad
  Jing Yang$^{4,5}$ \quad
  Nedjma Ousidhoum$^{6}$ \\
  \email amir@cis.lmu.de, nafiseh.nikeghbal@tum.de \\\\
  \addr $^{1}$LMU Munich  \\
  $^{2}$Technical University of Munich \\
  $^{3}$Munich Center for Machine Learning \\
  $^{4}$University of Southern California\\  $^{5}$Paper Copilot \\
  $^{6}$Cardiff University \\
}

\def\month{MM}  % Insert correct month for camera-ready version
\def\year{YYYY} % Insert correct year for camera-ready version
\def\openreview{\url{https://openreview.net/forum?id=XXXX}} % Insert correct link to OpenReview for camera-ready version

\begin{document}

\maketitle

\begin{abstract}

Peer review is a cornerstone of scientific publishing, including at premier machine learning conferences such as ICLR. 
As submission volumes increase, understanding the nature and dynamics of the review process is crucial for improving its efficiency, effectiveness, and the quality of published papers.
We present a large-scale analysis of the ICLR 2024 and 2025 peer review processes, focusing on before- and after-rebuttal scores and reviewer–author interactions. We examine review scores, author–reviewer engagement, temporal patterns in review submissions, and co-reviewer influence effects. Combining quantitative analyses with LLM-based categorization of review texts and rebuttal discussions, we identify common strengths and weaknesses for each rating group, as well as trends in rebuttal strategies that are most strongly associated with score changes. Our findings show that initial scores and the ratings of co-reviewers are the strongest predictors of score changes during the rebuttal, pointing to a degree of reviewer influence. Rebuttals play a valuable role in improving outcomes for borderline papers, where thoughtful author responses can meaningfully shift reviewer perspectives. 
More broadly, our study offers evidence-based insights to improve the peer review process, guiding authors on effective rebuttal strategies and helping the community design fairer and more efficient review processes. Our code and score changes data are available at \href{https://github.com/papercopilot/iclr-insights}{\path{github.com/papercopilot/iclr-insights}}.
\end{abstract}

\section{Introduction}

Peer review has been central to scientific publishing since 1665~\citep{kachooei2022peer}, with formal oversight beginning in 1752~\citep{kronick1990peer}. The process has long been regarded as a cornerstone of academic integrity~\citep{shah2022challenges, price2017computational}, but faces increasing challenges due to rapid scientific growth. In computer science, for example, conference publications carry disproportionate weight compared to journals~\citep{vrettas2015conferences, tomkins2017reviewer, kim2019author, meho2019using}.\footnote{Top-tier venues such as NeurIPS, ICML, and ICLR now receive tens of thousands of submissions annually, with ICLR 2026 reviewing over 19,000 individual papers.} Both academic pressures, such as ``publish or perish,''~\citep{van2012intended, grimes2018modelling} and technical advances, including LLM-assistance~\citep{liang2024mapping}, have contributed to increased submission numbers.

Rebuttals, or formal responses from authors to reviewers, are a routine part of conference peer review, particularly in machine learning (ML) conferences. They allow authors to clarify misunderstandings, provide additional evidence, and support for author-reviewer engagement. While rebuttals may improve paper quality, prior research suggests that author responses generally have only a marginal impact on final scores~\citep{gao-etal-2019-rebuttal}.
Building on these observations, and considering the additional effort required from authors, reviewers, and meta-reviewers, it is important to understand: 1) what proportion of papers are affected by rebuttal score changes, 2) when a conference should implement a rebuttal stage, 3) the association between rebuttals and score changes, 4) the association between reviewer comments and ratings, and 5) how authors can maximize the impact of their rebuttals. In this work, we investigate these questions to provide a clearer understanding of rebuttal effectiveness in peer review. To this end, we use a large-scale open dataset collected using the OpenReview API covering ICLR 2024 and 2025. We supplement this dataset with previous review scores---\ie the original scores that were later overwritten by updated scores---to analyze how rebuttals influence score changes (see our dataset details in \tabref{dataset}). We examine not only the frequency of rebuttal score changes but also temporal patterns of reviewer and author activity , as well as the relationship between textual content and the changes in scores after rebuttals. Our analysis leverages both traditional statistical methods and LLM-based techniques to identify what factors---such as evidence-backed clarification, reviewer engagement, and deviations from co-reviewers---contribute the most to effective rebuttals. Our analyses yield several consistent insights into the dynamics of ICLR peer review and rebuttals: 
\begin{enumerate}[nolistsep,noitemsep,leftmargin=*]
\item Rebuttals primarily affect borderline papers, with score changes concentrated in the mid-range (5–6), where even small shifts can change outcomes. We estimate that rebuttals influence roughly one in five top papers (based on the acceptance threshold) at ICLR through rank improvements. 
\item Late reviewer submissions may correlate with slightly higher scores, though the effect is minimal. Rebuttals submitted in the middle of the rebuttal period may be the most effective, while very early or last-minute rebuttals have less impact on engagement and score changes.  
\item Reviewer disagreements decrease by roughly 9–10\% after rebuttals, with the strongest convergence observed for high-quality papers (oral/spotlight) and minimal effect for low-scoring or rejected papers. 
\item When comparing common strengths and weaknesses across rating categories, low-rated papers tend to be criticized for writing, experimental, and methodological flaws, while high-rated papers are recognized for novelty and methodological soundness.
\item Initial scores and other reviewers' scores strongly predict rebuttal score changes, but rebuttals that provide evidence-backed clarifications and avoid generic or vague defenses are more likely to result in positive score changes.
\end{enumerate}

\input{tables/data_stat}

\section{Corpus}

ICLR is one of the few venues that fully publishes its peer review process~\citep{yang2025position, wang2023have}, licensed under CC BY 4.0\footnote{The license metadata for each note from authors and reviewers in ICLR OpenReview is under CC BY 4.0.} and hosted on OpenReview~\citep{soergel2013open}. All stages of the review process---including reviews, author responses, and final decisions---are publicly accessible, even for papers that were ultimately rejected. This transparency ensures that studies on peer review using ICLR data are not subject to biases such as survivorship~\citep{brown1992survivorship}.

Each submission is typically reviewed by three to five reviewers, who provide structured written feedback along with numerical scores. The review format includes a summary of the paper, identified strengths and weaknesses, and questions for the authors. Numerical scores are provided for soundness, presentation, contribution, confidence, and overall rating. The overall rating for ICLR 2024 and 2025 is a discrete number in $\{1, 3, 5, 6, 8, 10\}$.%
\footnote{The overall rating for ICLR 2026, however, is mapped to $\{0, 2, 4, 6, 8, 10\}$.}
After the initial reviews are released, authors can submit a rebuttal, and reviewers are encouraged to read them, engage in discussions, and revise their assessments accordingly, before the final decisions are made. Our corpus consists of data from ICLR 2024 and 2025, covering the full review cycle.
To gather this dataset, we used the OpenReview API, which provides structured access to all submission-related records, including reviews, rebuttals, decisions, and submission metadata. Although OpenReview retains archived versions, it only displays the final versions\footnote{Except for the abstract, PDF submission, and supplementary materials, whose archived versions are also publicly available.}, meaning that preliminary reviews submitted during the rebuttal phase are overwritten. To track score changes accurately, data must therefore be collected both immediately after the initial reviews and again after the rebuttal period ends.
We archived the dataset at the time of review releases both in 2024 and 2025, containing both before- and after-rebuttal scores, enabling the analysis of how reviewer assessments change. We release the data through the Paper Copilot platform~\citep{yang2025position, yang2025paper} and our GitHub repository.\footnote{\url{https://github.com/papercopilot/iclr-insights}}
Table~\ref{tab:dataset} provides basic statistics of our ICLR 2024 and 2025 corpus. The dataset includes over 19,000 papers and over 74,000 reviews. Due to technical issues at the time of review release, some papers (less than 5\%) had no recorded scores before the rebuttal; therefore, the number of score change records is less than the total number of reviews.
Figure~\ref{fig:paper_avg_rating_after_rebuttal} shows the distribution of submitted papers based on their average review ratings (after rebuttal) and their final decision outcomes.

\input{figures/banner}

\section{Empirical Analysis of Review and Rebuttal Dynamics}

\input{tables/increase_decrease_keep}

\subsection{Effect of Rebuttals}

\textbf{RQ: How many papers or reviews are affected by the rebuttal process?} Table~\ref{tab:increase_decrease_keep} shows paper- and review-level statistics for ICLR 2024 and 2025, grouped by whether the overall rating (per paper) or review score (per review) increased, decreased, or remained unchanged after rebuttal.
In both years, most review scores remained unchanged (2024: 81\%, 2025: 75\%), followed by increases (2024: 17\%, 2025: 23\%), with decreases being the least frequent outcome (2024: 1\%, 2025: 1\%). Among papers with increased scores, 42\% in 2024 and 44\% in 2025 were still not accepted despite improved reviewer assessments. However, papers with increased scores were far more likely to be accepted (57.6\% and 55.7\%) than those with unchanged (12.4\% and 7.8\%) or decreased (6.4\% and 8.0\%) scores.
Most score updates occurred in borderline cases, with the most frequent changes in this order: 5 → 6, 6 → 8, and 3 → 5, as shown in the \figref{heatmap_changes_comparison}.
Reviews with increased scores show higher engagement, including more conversation turns and stronger author–reviewer participation than in the \textit{keep} or \textit{decrease} score cases, indicating that active round-trip rebuttal discussions are often correlated with positive score changes. Interestingly, decreases also occur in cases with high conversation turns, showing greater variance than in the \textit{increase} category, suggesting that both lack of reviewer responses and active discussions can also be correlated with score reductions.

\input{figures/fg_heatmap_change_rating}

\textbf{RQ: Does score change affect paper ranking (proxy for acceptance)?}
If all review scores shift by a constant amount, paper rankings remain unchanged, making the effect of score changes neutral.
We show in \figref{displaced_top_spot} the percentage of papers displaced from the top ranks after rebuttal-induced score changes, measured as a function of the proportion of top papers considered. The displacement is most evident among the highest-ranked papers: in both years, over 40\% of papers in the top 5\% before rebuttal are replaced after score updates. This may occur because authors of papers potentially assured of acceptance at the top have little incentive to pursue higher scores, whereas those with lower---but still acceptable---scores may actively seek improvements. This results in large shifts within the top-ranked.
The overall patterns for 2024 and 2025 are consistent, with slightly higher displacement in 2025 across most thresholds. These results highlight that even small score updates during rebuttal can substantially affect the relative ranking of papers, particularly among top-ranked submissions where acceptance decisions are most competitive.
For ICLR 2024 and 2025, the acceptance rates are around 30\% (30.81\%, and 31.75\% respectively). As shown in \figref{displaced_top_spot}, nearly 20\% of papers lost their position in the top 30\% after rebuttal. This displacement drops to about half when the acceptance rate is set at 50\%. This indicates that rebuttals may matter more when acceptance rates are low, whereas for conferences (or workshops) with higher acceptance rates, rebuttals may have limited impact. Note that ranking alone does not determine acceptance, as other factors---such as meta-review assessments based on paper quality, topic interest, or the quality of reviews---can also determine the fate of a paper. Here, we use ranking as a proxy for acceptance and do not consider other positive effects of rebuttals that may exist, such as improvements in quality or changes in a paper's ultimate outcome. However, the impact of these factors are difficult to measure, as we cannot determine whether improvements in a revised paper are due to the rebuttal itself, the review, or other aspects.

\input{figures/time_order_2025_2024}

\subsection{Temporal Dynamics of Reviews and Rebuttals}\seclabel{on-time}
\textbf{RQ: Does Reviewer 2 exist?}
There is an inside joke in the ML (and other) communities about reviewer ordering; for example, Reviewer 2 (or 3) symbolizes the peer reviewer who writes vague or unhelpful reviews, assigns low scores, and refuses to budge during the rebuttal~\citep{watling2021don, worsham2022empirical, jin2023not, kinnear2025reviewer, tardy2018we, lundy2022praise}. Some studies have investigated whether such reviewers truly exist or if the perception arises solely from their assigned numbers~\citep{peterson2020dear}.
When examining the order of review submissions based on their creation timestamps, we find that the review submitted first appears last in the public display order; consequently, the most recently submitted review is shown first to all viewers.
Most of the ICLR papers had four reviews (2024: 65\%, 2025: 67\%). In this part, we only consider papers with four reviews and assign review numbers 1 to 4.
As shown in \figref{reviewer_time_order_bar_2025_position}, the distribution of scores, regardless of the papers they are assigned to, appears similar across reviewers. However, Reviewers 4, 3, 2, and 1, in order, tend to show increasing generosity-, which means that the reviewer who submits last (Reviewer 1) tends to submit higher scores. The result is the same for ICLR 2024 in \figref{reviewer_time_order_bar_2024_position}.
Even though the reviews in \figref{reviewer_time_order_bar_2025_position} appear relatively similar, ordering the review scores per paper reveals a different picture in \figref{reviewer_time_order_bar_2025_minmax}. Every paper consistently receives both high and low scores. %The Reviewer 2 joke is largely a product of confirmatory bias~\citep{nickerson1998confirmation}: when a poorly given score happens to coincide with Reviewer 2, it is reported and amplifies the community belief~\citep{zhu2025survivors}, even though low scores occur across all reviewer positions. 
\figref{reviewer_time_order_bar_2025_minmax} also shows that the most frequently assigned low scores are 3, middle scores are 5, and the highest scores are 6.

\input{figures/author_reviewer_activity}

\textbf{RQ: When are reviews and rebuttals typically submitted during the peer review process?}
\figref{author_reviewer_activity_2025} shows the daily activity of reviewers and authors in ICLR 2025, measured by the number of submitted messages across the review and rebuttal phases. ICLR deadlines use AoE, but we use UTC here, in conformity with the default on OpenReview.
During the review-writing phase (October 15 – November 4), activity is exclusively from reviewers, peaking on November 3–4, immediately before the review deadline. This indicates that reviewer activity is highly concentrated around the deadline. After the submission of reviews, activity drops for emergency reviews until the reviews are released, marking the start of the rebuttal phase on November 12.
In the early days of the rebuttal period, author contributions are limited yet more prominent compared to reviewers. Reviewer and author activities rise later, peaking on November 25–26 for reviewers, near the official rebuttal deadline. Even after the rebuttal period is extended, overall activity remains low.
Interestingly, a similar pattern occurred in ICLR 2024 (\figref{author_reviewer_activity_2024}), but with a difference: reviewers were the most active on the last day, whereas in 2025 authors were more active---likely due to a message from the committee or area chairs prompting reviewer action. This pattern shows that the activities are mostly shaped by the deadlines of each group rather than by steady interaction.

\input{figures/author_activity_first_message}

\textbf{RQ: When is the best time for authors to start the rebuttal?} 
The rebuttal starts with the first message; usually, authors prepare their entire rebuttal in this initial interaction. \figref{author_activity_first_message_2025} shows daily author activities in ICLR 2025. We color the first messages based on whether the reviewer later changed their score. Numbers above each bar indicate the percentage of first messages that led to a later review score increase. As expected, messages submitted late---after or near the original rebuttal deadline---are less often associated with score increases. Interestingly, messages submitted very early were also less successful, which may be attributed to the rebuttal being rushed or to the papers having received very low scores. During the period November 18–24, nearly one-third of first messages later led to a review score increase. The same pattern, with some differences, also holds for ICLR 2024 (\figref{author_activity_first_message_2024}).

\subsection{Co-Reviewers Influence}
We use the term ``co-reviewers influence'' to refer to the incentive for reviewers to update their scores and reach a consensus. Such consensus is often explicitly encouraged by area chairs, particularly when there is high deviation among review scores and new reviewers cannot be assigned. Co-reviewers influence has also been described in other contexts as peer pressure~\citep{gao-etal-2019-rebuttal}, herd behavior~\citep{banerjee1992simple}, conformity bias~\citep{buechel2015opinion}, or balance between reviewers~\citep{huang2023makes}; we do not differentiate between these forms. Since we do not have access to controlled experimental data (i.e., a control group of reviewers without exposure to others’ reviews), our analysis cannot fully attribute the observed convergence of scores solely to co-reviewers influence. We formalize co-reviewers influence as the extent to which pairs of scores move closer together. This can be computed per review pair or per paper across all reviewer pairs. For reviewers $i$ and $j$ with scores $s_i$ and $s_j$, disagreement is $|s_i - s_j|$. For a paper with $n$ reviewers, average disagreement is \(\frac{1}{\binom{n}{2}} \sum_{i < j} |s_i - s_j|\).

We compute reviewer disagreement before and after rebuttals and calculate the normalized difference $(\text{after} - \text{before}) / 9$, where 9 is the score range. A larger decrease indicates a stronger co-reviewers influence.

\textbf{RQ: From when can we expect co-reviewers influence to take effect?} Influence can occur as soon as reviews are released. When examining the last modification timestamp of the reviews upon release; initially, each timestamp is set to the release time. We observe that many changes---mostly increases in scores---occur within the first hour of release (see \figref{last_modification_date}).

\input{figures/last_modification_date}

\input{tables/reviewer_venue_distance}

\textbf{RQ: Do reviewers' scores diverge or converge?}
\tabref{reviewer_venue_distance} presents the average reviewer disagreement across different submission categories and in total for ICLR 2024 and 2025. Disagreements are reported for the before and after rebuttal, with both absolute ($\Delta$) and relative (Rel.\ $\Delta$\%) changes also shown.
The results demonstrate a consistent reduction in disagreements after rebuttal in both years, indicating that reviewer evaluations become more balanced and aligned during this phase. The effect, however, is not uniform across categories. For spotlight and oral submissions, reductions are largest: 29\% \& 26\% for spotlight and 41\% \& 48\% for oral in 2024 and 2025, respectively.
This indicates that potentially high-quality submissions are more strongly affected by co-reviewers influence, leading to greater reviewer convergence. Poster submissions show more moderate but consistent decreases. By contrast, rejected and withdrawn submissions show only small reductions (generally below 7\%), indicating limited rebuttal influence when scores are low. Overall, average distances drop by 9-10\%. This shows that rebuttals can act as a balancing step, consistently reducing reviewer divergence, with the effect most pronounced in top-tier categories, which likely would have been accepted at least as a poster even without this change.

\section{LLM-Based Analysis of Review and Rebuttal Dynamics}

\subsection{Reviewer Evaluation Patterns}\seclabel{llm-review}

\textbf{RQ: What are the most common strengths and weaknesses per paper and overall rating group?} We show the common strengths and weaknesses per rating group in Appendix \figsref{taxonomy_weakness}{taxonomy_strength}. Our analysis reveals a clear gradient: as ratings increase, the perception of weaknesses decreases, while the recognition of strengths grows. Low-rated papers are dominated by writing flaws, such as unclear wording, and experimental flaws, such as weak baselines, whereas high-rated papers are highlighted for novelty, such as original ideas, and methodology, such as elegant and efficient models. Mid-range ratings reflect a mixed evaluation, where reviewers balance both weaknesses and strengths, often emphasizing promising contributions that are still underdeveloped.

We extract this information by prompting GPT-4o to identify strengths and weaknesses in each review. To ensure consistency, we provide GPT-4o with a structured taxonomy of high-level categories and subcategories that the model must use when labeling the text. For weaknesses, the categories cover aspects such as Novelty \& Contribution (e.g., lack of originality, incremental improvement), Motivation (e.g., weak justification of the problem), Methodology \& Technical Soundness (e.g., unrealistic assumptions, unclear algorithmic description), Experiments \& Evaluation (e.g., insufficient baselines, missing ablations), Results (e.g., marginal gains), Data (e.g., poor data quality), Writing \& Presentation (e.g., unclear wording), Broader Impact \& Ethics, Related Work, and Venue Fit. The strength taxonomy mirrors this structure, including subcategories such as high novelty, strong theoretical justification, broad and realistic experimental coverage, substantial gains, high-quality data, and clear presentation. 
The categories were iteratively refined with input from two academic experts, based on over 100 review statements from ICLR 2024–2025 and the seed weakness categories of \citet{gao-etal-2019-rebuttal} (see \Appref{design-prompts}). 
The full prompt templates and complete category definitions—with all subcategories—are provided in the Appendix \figsref{weakness_prompt}{strength_prompt}. To reduce computational load, we select a sample of 4,000 papers from both ICLR 2024 and 2025, evenly distributed across different paper outcomes during the rebuttal phase.

\input{figures/fig_review_scores_and_heatmap}

\textbf{RQ: Which strengths and weaknesses influence the overall score the most?}
We show the top categories associated with strengths and weaknesses in \tabref{feature_importance}. To compute these categories, we use word counts as a strong feature (see next RQ) and category appearance vectors (from the previous RQ) for `strengths' and `weaknesses' as additional features.
The output labels represent the overall rating (before rebuttal) for each review. They are discrete and carry meaningful information, so we treat them as distinct classes and perform classification.
For scores 1 and 3, we assign the category \textit{Low}, and similarly assign the category \textit{High} to scores 8 and 10 due to their low frequency of occurrence. The borderline scores 5 (Low-B) and 6 (High-B) are treated as separate classes.
We use multinomial logistic regression, splitting the data into training (80\%), validation (10\%), and test (10\%) sets, and normalize all features using a scaler fitted on the training data. We perform a grid search over class weights based on the validation set. We choose multinomial logistic regression for its interpretability as it allows us to identify important features. When using all features, the macro-f1 score is 0.49, compared to 0.43 when only using word counts. The top-ranked features, according to their average absolute coefficient (Avg. $|$Coef.$|$), are shown in \tabref{feature_importance}. 
A closer look at the subcategories  of the top two features (see \figref{taxonomy_weakness}) shows that for `Novelty \& Contribution', the most frequent issues are overlap with prior work, lack of originality, and lack of clear contribution. For `Experiments \& Evaluation', the most common concerns are insufficient or weak baselines, too few datasets or limited domain, missing ablation tests, and reproducibility issues.

\textbf{RQ: What further insights can be derived from the relationship between review scores and textual features, such as word count?}
\figref{correlation_heatmap} shows the correlations between score-based criteria (SC) and word count features (WC). All non-zero correlations have $p$-values below $0.001$.
We find that the overall rating correlates most strongly with contribution, then soundness.
The overall rating shows a positive correlation with the word count of the strengths section and a negative correlation with the word count of the weaknesses section. Interestingly, reviews with higher confidence scores tend to submit longer weaknesses sections and are associated with lower overall ratings.
\tabref{stats_per_category} supports these findings by breaking down average scores and word counts across submission categories. Accepted papers (oral, spotlight, and poster) generally receive higher scores and longer reviewer text in the summary, strengths, and questions sections, whereas rejected submissions (desk-rejected, withdrawn, and rejected) receive lower scores and longer weaknesses sections. This pattern suggests that reviewers provide more strengths in high-quality papers, and more extensive critical or constructive feedback for weaker papers.

\input{tables/stats_per_category}
\input{tables/score_changes_summary}

\subsection{Common Rebuttal Strategies and Outcomes}

\textbf{RQ: Which rebuttal strategies are associated with changes in reviewer scores?}  
To answer this question, we only consider data for papers whose authors participated in the rebuttal, as a lack of author participation could result in reviewers also not participating.
Here, we only consider the overall rating score change, as most reviews (71\% of cases, see \tabref{score_changes_summary}) reflect changes only in the overall score and do not update other aspects, such as soundness. %We examine strategies employed during the rebuttal phase to convince the reviewer to change their scores. 
We aim to identify which strategies, used to convince the reviewer to change their scores, are more likely to lead to success or failure. Hence, we prompt GPT-4 to annotate strategies based on the given categories for the same batch of sampled papers in \secref{llm-review}. The categories were iteratively
refined with input from two academic experts, based on over 100 review statements from ICLR 2024–2025
and the seed starting with the strategies suggested by~\citet{noble2017ten, kennard-etal-2022-disapere, huang2023makes, gao-etal-2019-rebuttal, li2025effective} (see \Appref{design-prompts}). The final categories are presented in Appendix ~\figref{argument_prompt}. We use multinomial logistic regression to predict score changes for three classes---``increase'', ``decrease'', and ``keep''---because it offers interpretability, allowing us to identify important features. The data is split into training (80\%), validation (10\%), and test (10\%) sets, and features are normalized using a scaler fitted on the training data. We perform a grid search over class weights based on performance on the validation set.

We show the results of training and testing on both three-class and two-class settings---following the approach suggested by \citet{huang2023makes}, who merge ``decrease'' and ``keep'' into a single group---in \tabref{score_change_prediction}. In the three-class setting, ``keep'' has the largest class size, while ``decrease'' has the smallest, and the macro F1-score for each class correlates with these sizes. The overall macro F1 is 0.52 for the three-class task and 0.71 for the two-class task. This prediction task is challenging because several factors, including paper quality and reviewer characteristics, are not considered in the model. We also show the top features and their coefficients, which indicate the importance of each strategy in the 3-class classification in \tabref{top_features_coef}. Some of the most important features are aspects over which the author has limited control in the rebuttal phase, such as the initial overall rating, the contribution and soundness scores, and the average of other reviewers' scores. A notable feature is reviewer engagement, which can lead to either an increase or a decrease in scores. Although authors can respond to engagement, the effect ultimately depends on the reviewers' willingness to participate. Other important features are more strategy-based. For example, a clearly ``evasive stance'' or a ``generic/vague defense'' strategy tends to only help reviewers maintain their original scores, while ``Bare agreement/disagreement'' and providing ``evidence-backed clarification'' can help increase the scores.

\input{tables/score_change_prediction}

\input{tables/rebuttal_strategy}

\section{Related Work}\seclabel{related-work}

The peer review process has been the focus of multiple research efforts on various NLP tasks~\citep{lin2023automated, drori2024human, kuznetsov2024can, staudinger-etal-2024-analysis}, including review and rebuttal analysis, among others (see \Appref{appendix-related-work} for more).

\textbf{Review Analysis.}
Prior work has examined the content and characteristics of reviews, including their quality and tone. For example, studies have measured review length and overall quality~\citep{geldsetzer2023prevalence}, evaluated politeness or harshness in peer reviews~\citep{verma-etal-2022-lack,bharti2024politepeer}, detected misinformed or deficient review points~\citep{ryu2025reviewscore, zhang2025reviewguard}, assessed the utility of reviews for authors~\citep{sadallah2025good}, explored argumentative perspectives to identify disagreements between reviewers and scoring discrepancies across submission versions~\citep{gao-etal-2019-rebuttal, chakraborty2020aspect}, and analyzed reviewer confidence in specific sections or aspects of a paper to derive insights into overall review quality. Since reviewers are typically asked to provide numerical scores alongside their textual feedback, some studies have focused on predicting scores or acceptance decisions from textual features~\citep{kang-etal-2018-dataset, fernandes2022between, fernandes2024enhancing}; for instance,~\citet{ghosal-etal-2019-deepsentipeer, ribeiro2021acceptance} applied sentiment analysis techniques to estimate acceptance likelihood based on review language.

In our work, we similarly focus on the content of reviews, but identify common categories of strengths and weaknesses, examine their relationship with different rating groups, and use them to model scoring patterns. The interpretability of this framework further enables us to determine which review attributes most strongly influence reviewer decisions.

\textbf{Rebuttal Analysis.}
Some peer review processes include a conversation between authors and reviewers, conducted to resolve misunderstandings, address comments, and provide additional experiments. The effects of this engagement can be reflected in the rating score and final version of a publication. Therefore, analyzing the differences between successive submission versions, as well as interactions between reviewers and authors and corresponding score changes, is relevant. Several datasets have been developed to study these phenomena~\citep{gao-etal-2019-rebuttal, hua-etal-2019-argument, cheng-etal-2020-ape, choudhary2021react, kennard-etal-2022-disapere,  huang2023makes, purkayastha-etal-2023-exploring, ruggeri-etal-2023-dataset, bharti2024politepeer}.

Among these, the most closely related papers are by~\citet{gao-etal-2019-rebuttal} and~\citet{huang2023makes}. The former introduced an open corpus from ACL 2018 and analyzed before- and after-rebuttal score changes, though their dataset may be biased since 69\% of authors withheld consent to share responses. The latter examined over 3,000 ICLR 2022 papers, treating rebuttals as social interactions between authors and reviewers and identifying common rebuttal strategies, but did not release the underlying data or score changes.

In our work, we extend this line of research by analyzing a larger corpus of ICLR submissions and exploring new angles, including when rebuttals are most effective, how timing is associated with their impact, and which factors, identified through LLMs, are associated with reviewer scores and rebuttal outcomes.\footnote{Our concurrent work~\citep{jung2025drives}, focuses on acceptance decisions and does not include experiments on score changes, which is the main focus of our study. In contrast, we analyze score dynamics and do not examine acceptance decisions, in order to avoid potential incentives for gaming or reverse-engineering the factors that influence paper acceptance.}

\section{Conclusion}

We aim to provide a systematic account of how rebuttals and reviewer dynamics shape outcomes in large-scale conference peer review. To this end, we analyze ICLR 2024 and 2025 review data, combining statistical methods with LLM-based categorization of review texts and rebuttal exchanges. Our study reveals that rebuttals primarily affect borderline papers, with approximately 20\% of accepted submissions likely benefiting from rebuttal-driven score increases.
We further find that initial reviewer scores and co-reviewer ratings are the strongest predictors of rebuttal score changes, indicating substantial peer influence. Reviewer disagreements narrow after rebuttals, especially for high-scoring paper submissions, while low-scoring ones remain largely unaffected. Textual analysis highlights that evidence-backed clarifications and precise responses are most strongly associated with positive changes, whereas vague or defensive rebuttals have little effect.
These results offer practical guidance for authors in crafting effective rebuttals and shed light on the role of reviewer interactions in shaping final outcomes. More broadly, they inform program chairs seeking to design review processes that balance fairness and efficiency amid increasing submission volumes.

\section*{Limitations}\seclabel{limitations}
1) Our findings should be interpreted with caution. Although our analyses reveal clear statistical patterns, the results should be interpreted as correlational and descriptive, not causal. 

2) While the findings of this paper are specific to ICLR, they provide valuable insights tailored to this venue, which is the primary focus of our study.

3)Although a small subset of the data may be missing---for example, in cases where the meta-review references an author–reviewer discussion that is not available or or due to a small portion of lost score change records from technical issues---the overall scale of our dataset ensures that these gaps are unlikely to meaningfully impact the findings. 

4) We rely in part on LLM-based categorization of review and rebuttal texts to identify strengths, weaknesses, and strategies and LLM judgments remain imperfect. However, we iteratively refined the categories and prompts with expert input and achieved relatively high pairwise agreement with human annotations. 

5) We predict review scores and rebuttal outcomes only to identify features most strongly correlated with them, not to set predictive performance benchmarks. Accordingly, we used an interpretable multinomial logistic regression model, and the results should be read as exploratory. 

6) We primarily analyze the review and rebuttal stages, without considering the meta-review stage, which may affect borderline papers. Since meta-reviews largely depend on the quality of the reviews and the perceived expertise of the reviewers, focusing on the review and rebuttal stages allows us to more clearly isolate the dynamics that directly drive score changes. We leave the meta-review analysis for future work.

\section*{Reproducibility Statement}

We do not publish any OpenReview data, as it is already publicly available and can be accessed at any time through the official OpenReview API to reproduce our results. We share the ICLR 2024 and 2025 score changes we obtained---which can no longer be accessed through standard means---under the same OpenReview license, CC BY 4.0. All experimental code is open source. The prompts are provided in the Appendix and will also be shared alongside the code. The exact GPT-4o model used in these experiments is \texttt{gpt-4o-2024-08-06}, with \texttt{top\_p=0} and \texttt{temperature=0}. The total computation cost for the experiments was \$500 in OpenAI credits.  

\section*{Ethics statement}
This study investigates the peer review and rebuttal dynamics of ICLR 2024 and 2025 using publicly available data from OpenReview, which is licensed under CC BY 4.0. We did not collect any private or confidential data. All reviews, rebuttals, and decisions are already openly accessible as part of ICLR’s transparent peer review policy. To respect the integrity of reviewers and authors, we focus on aggregated analyses and refrain from attributing any results to specific individuals.

Although peer reviews may contain strong critiques, disagreements, or subjective opinions, our analysis treats them as research artifacts for understanding trends in review and rebuttal processes, not for evaluating individual reviewers or authors. Although we acknowledge that automated text analysis with LLMs may introduce limitations or biases in categorization, these methods were only used to identify general patterns of strengths, weaknesses, and rebuttal strategies. Human experts guided the design of annotation categories to ensure the extracted insights were meaningful and appropriate.

The purpose of this work is to provide constructive insights for authors, reviewers, and program chairs, with the overarching goal of improving fairness, efficiency, and transparency in scientific peer review. We refrain from making any causal claims, as our analyses are based on observational data and cannot disentangle causal effects from underlying confounding factors. Consequently, all reported relationships should be interpreted as correlational rather than causal. To facilitate transparency and reproducibility, we release our code, prompts, and derived score-change data under the same CC BY 4.0 license as the original OpenReview records.

\bibliography{tmlr}
\bibliographystyle{tmlr}

\appendix

\section{Appendix}

\subsection{Additional Related Work}\seclabel{appendix-related-work}

In NLP research on peer review, beyond analyzing reviews and rebuttals, several other main tasks have been studied, including reviewer assignment~\citep{stelmakh2021catch, jecmen2023dataset, jecmen2025on, stelmakh2025a}, 
review generation~\citep{wang-etal-2020-reviewrobot, yuan2022can, liu2023reviewergpt, szumega2023open, zhou-etal-2024-llm, zhang2025re}, 
meta-review generation~\citep{shen-etal-2022-mred, wu2022incorporating, li-etal-2023-summarizing, lin2023moprd, zeng2023meta, sun2024metawriter}, 
guided skimming~\citep{dycke-etal-2023-nlpeer}, and citation prediction~\citep{li-etal-2019-neural, plank2019citetracked}.  
More recently, new tasks have emerged, including the use of LLMs as tools for reviewing, which raises concerns about quality and fairness~\citep{choi2025position, sun2025openreview, zhu2025your, thakkar2025can, wei2025ai, li2025unveiling, li2025llm}. These developments have sparked debates around the so-called ``AI conference peer review crisis''~\citep{kim2025position, chen2025position}, highlighting the need to rethink peer review management in computer science, including proposals to establish dedicated tracks for refutations and critiques in ML conferences~\citep{schaeffer2025position}.

\subsection{Design of Prompts}\seclabel{design-prompts}

\textbf{Weakness/Strength Categories.}
We adopt weakness categories from \citet{gao-etal-2019-rebuttal} as a starting point and use them, to also define corresponding strength categories, in order to classify both weaknesses and strengths. To refine the categories, two academic experts (authors of this paper; compensated according to their employment contract) independently reviewed 20 papers selected from different decision outcomes (reject, oral, spotlight, etc.) and annotated the weaknesses and strengths in their reviews. Weaknesses and strengths were either assigned to existing categories or, when necessary, placed into newly proposed categories or subcategories.
After this initial round, the authors met to compare results and reached consensus on a preliminary taxonomy. The goals of this meeting were (a) to merge overlapping categories and (b) to validate the newly proposed categories. To test the stability of the taxonomy, the authors then reviewed an additional 10 papers (later also used as a validation set) and compared their results. Since no new categories or subcategories emerged, the taxonomy was finalized.

Based on the outcomes of these two rounds, we established a stable set of categories and subcategories. Using this refined taxonomy, we designed prompts for GPT-4o to classify reviews at scale. We experimented with different prompt designs on the validation set, enabling comparison between GPT-4o outputs and our manual annotations. We then selected the prompt that achieved the highest alignment with the manual labels (84\% pairwise human agreement) and applied it to a larger set of review statements (see the weakness prompt in \figref{weakness_prompt} and the strength prompt in \figref{strength_prompt}).

\figref{taxonomy_weakness} and \figref{taxonomy_strength} present the main weaknesses and strengths highlighted by reviewers at ICLR 2024 and 2025, organized by paper rating scores before rebuttal. For each rating group (1, 3, 5, 6, 8, and 10), the most common categories of criticism and praise are identified, along with their top three subcategories. Low-rated papers are often criticized for unclear writing, weak baselines, limited datasets, poor unclear algorithmic description, while high-rated papers are recognized for novelty, methodological soundness, efficient model or approach, and clear presentation.

\textbf{Strategy Categories.}
We adopt strategy categories from \citet{noble2017ten, gao-etal-2019-rebuttal, kennard-etal-2022-disapere, huang2023makes, li2025effective} as a starting point. Like the weakness and strength category design,  the same two academic experts independently selected 10 papers from each category (increase, decrease, keep) to develop more fine-grained strategies by defining subcategories. They further discussed these strategies in a meeting and finalized a stable set of categories and subcategories. Using this refined taxonomy, we designed prompts for GPT-4o to classify reviews at scale. We validated different prompt designs against manual annotations of additional 10 papers and selected the one with the highest alignment (81\% pairwise human agreement). Our taxonomy consists of three main fields: \textit{coverage}, \textit{stance}, and \textit{strategy}. Coverage indicates whether the author addresses a reviewer’s concern. Stance reflects the author's position---agreement or disagreement with the reviewer's point---thereby revealing alignment or conflict. Strategy characterizes how the response is formulated, distinguishing evidence-backed clarifications from vague or generic defenses, which allows us to assess the substance and persuasiveness of rebuttals. Each of these fields contains a set of subcategories, which are detailed in the \figref{argument_prompt}.

\input{figures/taxonomy_weakness}
\input{figures/taxonomy_strength}
\input{figures/weakness_prompt}
\input{figures/strength_prompt}
\input{figures/argument_prompt}

\end{document}

%% file: math_commands.tex
\usepackage{amsmath,amsfonts,bm}

\def\figref#1{figure~\ref{#1}}
\def\secref#1{section~\ref{#1}}
\def\eqref#1{equation~\ref{#1}}
\def\1{\bm{1}}

\DeclareMathAlphabet{\mathsfit}{\encodingdefault}{\sfdefault}{m}{sl}
\SetMathAlphabet{\mathsfit}{bold}{\encodingdefault}{\sfdefault}{bx}{n}

%% file: boxs.tex
\tcbset{
  sbox/.style={
    colback=blue!5!white,
    colframe=blue!50!black,
    fonttitle=\bfseries,
    title=Reviewer–Author Stance \& Strategy Annotation Prompt,
    boxrule=0.5mm,
    arc=2mm,
    top=2mm,
    bottom=2mm,
    left=2mm,
    right=2mm,
    fontupper=\footnotesize	
  }
}

\tcbset{
  wbox/.style={
    colback=purple!5!white,
    colframe=purple!50!black,
    fonttitle=\bfseries,
    title=Weakness Annotation Prompt,
    boxrule=0.5mm,
    arc=2mm,
    top=2mm,
    bottom=2mm,
    left=2mm,
    right=2mm,
    fontupper=\footnotesize	
  }
}

\tcbset{
  gbox/.style={
    colback=green!5!white,
    colframe=green!50!black,
    fonttitle=\bfseries,
    title=Strength Annotation Prompt,
    boxrule=0.5mm,
    arc=2mm,
    top=2mm,
    bottom=2mm,
    left=2mm,
    right=2mm,
    fontupper=\footnotesize	
  }
}

\tcbset{
  pbox/.style={
    colback=yellow!10!white,
    colframe=yellow!50!black,
    fonttitle=\bfseries,
    title=Politeness Annotation Prompt,
    boxrule=0.5mm,
    arc=2mm,
    top=2mm,
    bottom=2mm,
    left=2mm,
    right=2mm,
    fontupper=\footnotesize	
  }
}

\tcbuselibrary{skins}

\definecolor{keycolor}{rgb}{0,0,0.6}
\definecolor{stringcolor}{rgb}{0.64,0.16,0.29}
\definecolor{valuecolor}{rgb}{0,0.5,0}

%% file: tables/data_stat.tex
\begin{table}[t]
\centering
\footnotesize
\caption{Statistics of papers, reviews, and score-change records in our ICLR 2024 and 2025 corpus.}
\label{tab:dataset}
\begin{tabular}{cccc}
\toprule 
\textbf{Year} & \textbf{\#Paper} & \textbf{\#Review} & \textbf{\#Score Change Record} \\
\midrule
\textbf{2025} & 11672 & 46748 & 46353 \\
\textbf{2024} & 7405  & 28028 & 26878 \\
\bottomrule
\end{tabular}
\end{table}

%% file: figures/banner.tex
\begin{figure}[t]
    \centering
    % Left figure
    \begin{minipage}[t]{0.48\linewidth}
        \centering
        \includegraphics[width=\linewidth]{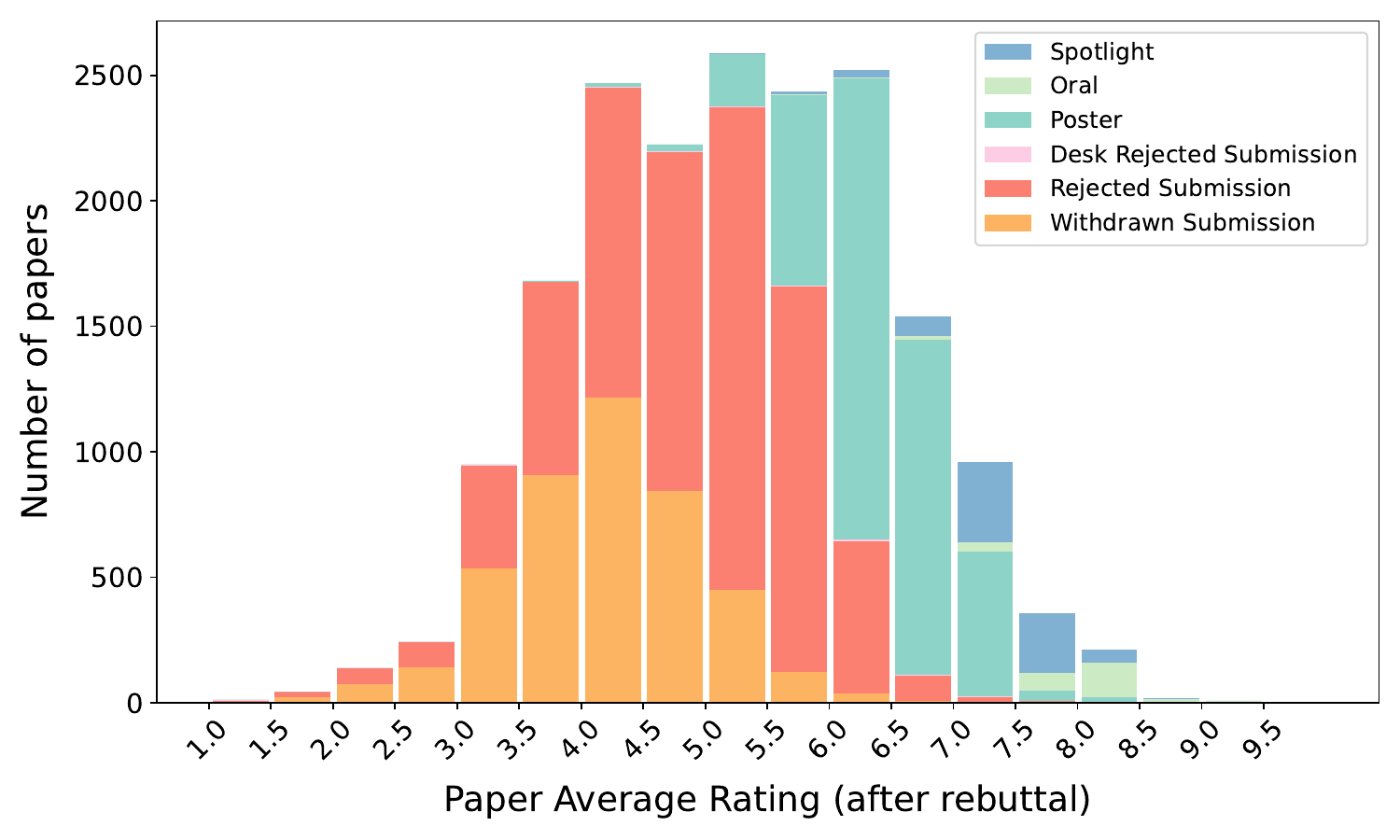}
        \caption{Distribution of submitted papers by average overall rating score and final decision.}
        \label{fig:paper_avg_rating_after_rebuttal}
    \end{minipage}%
    \hfill
    % Right figure
    \begin{minipage}[t]{0.41\linewidth}
        \centering
        \includegraphics[width=\linewidth]{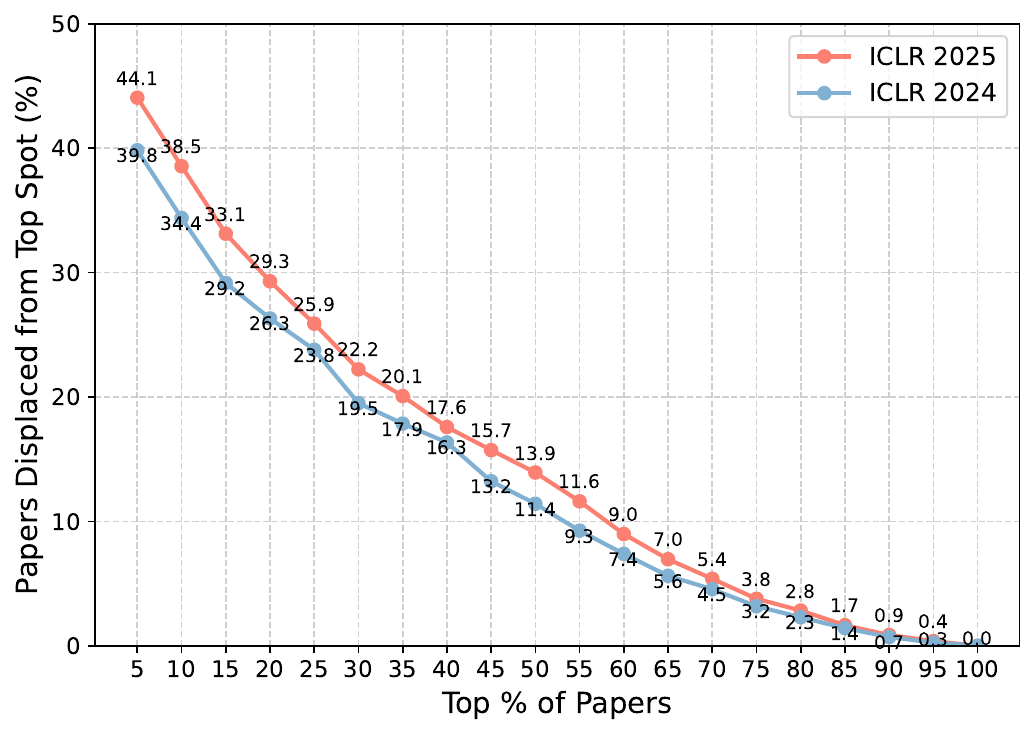}
        \caption{Percentage of papers displaced in top \% threshold in 2024 and 2025 (before → after rebuttal).}
        \label{fig:displaced_top_spot}
    \end{minipage}
\end{figure}

%% file: tables/increase_decrease_keep.tex
\begin{table}[h]
\caption{Paper- and review-level statistics for our ICLR 2024–2025 corpus, showing score changes ($\Delta$) after rebuttal, acceptance rates (Acpt.\%), and discussion metrics: author/reviewer participation percentage (Part\%) and conversation turns between authors and reviewers (ConvTurn).}
\label{tab:increase_decrease_keep}
\resizebox{0.91\textwidth}{!}{%
\begin{minipage}{0.48\textwidth}
\centering
\footnotesize
\renewcommand{\arraystretch}{1.2}
\vspace{4pt}
\begin{tabular}{lcccc|}
\toprule
\textbf{Year} & \textbf{Type} & \multicolumn{3}{c}{\textbf{Per Paper}} \\
\cmidrule(lr){3-5} 
 & & {\#Paper} & {Acpt.\%} & {$\Delta_\mathrm{Rating}$} \\
\hline
\multirow{4}{*}{\textbf{2025}} 
 & {Increase} & 5807 & 55.7 & \cellcolor{blue!15} 5.21 → 5.97 \\
 & {Decrease} & 377  & 8.0  & \cellcolor{tabred!40} 4.88 → 4.41 \\
 & {Keep}     & 5247 & 7.8  & \cellcolor{gray!15} 4.30 → 4.30 \\
 & {Total}    & 11431& 32.1 & \cellcolor{blue!15} 4.78 → 5.15 \\
\hline
\multirow{4}{*}{\textbf{2024}}
 & {Increase} & 2930 & 57.6 & \cellcolor{blue!15} 5.31 → 6.01 \\
 & {Decrease} & 251  & 6.4  & \cellcolor{tabred!40} 4.91 → 4.44 \\
 & {Keep}     & 3792 & 12.4 & \cellcolor{gray!15} 4.51 → 4.51 \\
 & {Total}    & 6973 & 31.2 & \cellcolor{blue!15} 4.86 → 5.14 \\
\bottomrule
\end{tabular}
\end{minipage}
% \hfill
\hspace{-10pt}
\begin{minipage}{0.48\textwidth}
\centering
\footnotesize
\renewcommand{\arraystretch}{1.2}
\vspace{4pt}
\begin{tabular}{lcccc}
\toprule
\multicolumn{5}{c}{\textbf{Per Review}} \\
\cmidrule(lr){1-5} 
{\#Reviews} & {ConvTurn} & {AuthPart\%} & {RevPart\%} &  {$\Delta_\mathrm{Rating}$} \\
\hline
10728 & 2.21 ± 0.83 & 95.65 & 86.88 & \cellcolor{blue!15} 4.64 → 6.34 \\
640   & 2.04 ± 1.13 & 83.75 & 66.41 & \cellcolor{tabred!40} 5.95 → 4.10 \\
34985 & 1.47 ± 0.68 & 65.92 & 39.07 & \cellcolor{gray!15} 4.81 → 4.81 \\
46353 & 1.65 ± 0.79 & 73.05 & 50.51 & \cellcolor{blue!15} 4.78 → 5.15 \\
\hline
4666  & 2.04 ± 0.75 & 99.79 & 80.73 & \cellcolor{blue!15} 4.58 → 6.30 \\
359   & 1.71 ± 0.89 & 90.25 & 50.7  & \cellcolor{tabred!40} 6.28 → 4.49 \\
21853 & 1.35 ± 0.57 & 75.26 & 30.65 & \cellcolor{gray!15} 4.91 → 4.91 \\
26878 & 1.47 ± 0.67 & 79.72 & 39.61 & \cellcolor{blue!15} 4.86 → 5.14 \\
\bottomrule
\end{tabular}
\end{minipage}
}
\end{table}

%% file: figures/fg_heatmap_change_rating.tex
\begin{figure}[t]
    \centering
\begin{subfigure}[t]{0.33\textwidth}  % slightly wider
    \centering
    \includegraphics[width=\textwidth]{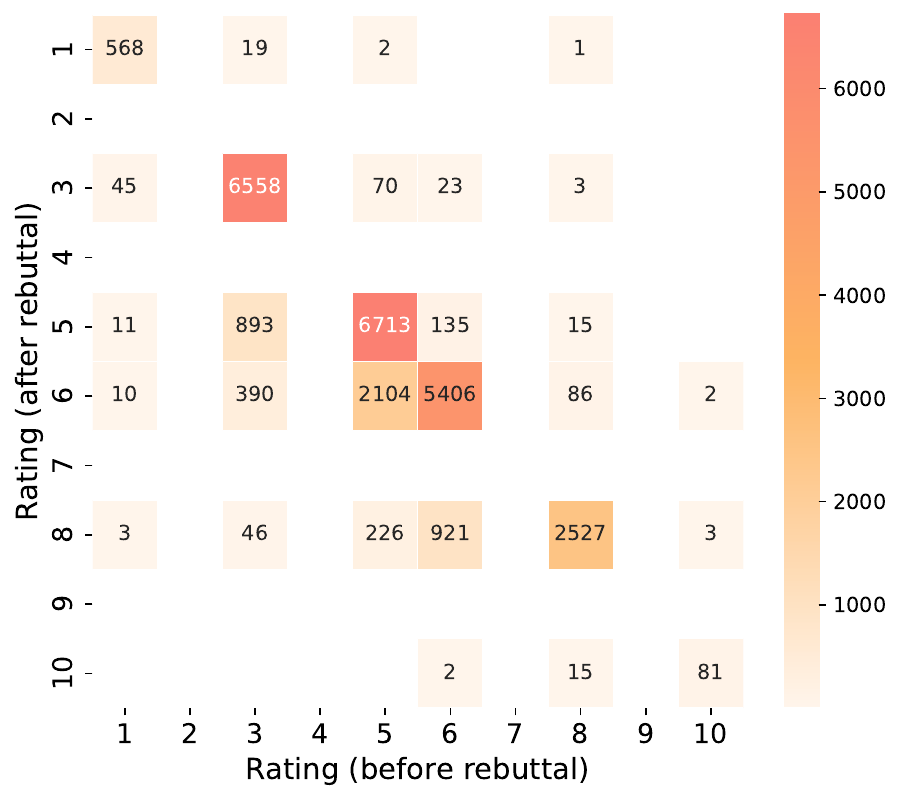}
    \caption{ICLR 2024}
    \label{fig:heatmap_changes_2024}
\end{subfigure}
\hspace{0.02\textwidth} % small horizontal space
\begin{subfigure}[t]{0.33\textwidth}
    \centering
    \includegraphics[width=\textwidth]{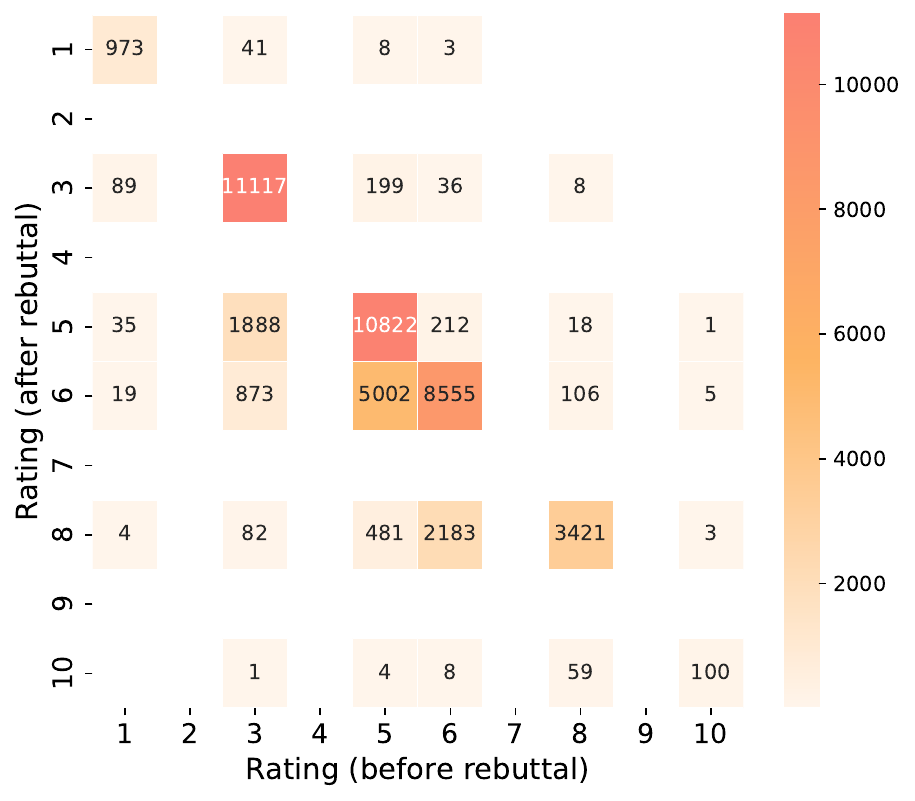}
    \caption{ICLR 2025}
    \label{fig:heatmap_changes_2025}
\end{subfigure}
    \caption{Change in rating scores from before (x-axis) to after (y-axis) the rebuttal.}
    \label{fig:heatmap_changes_comparison}
\end{figure}

%% file: figures/time_order_2025_2024.tex
\begin{figure}[t]
    \centering
    \begin{subfigure}[t]{0.23\textwidth}
        \centering
        \includegraphics[width=\textwidth]{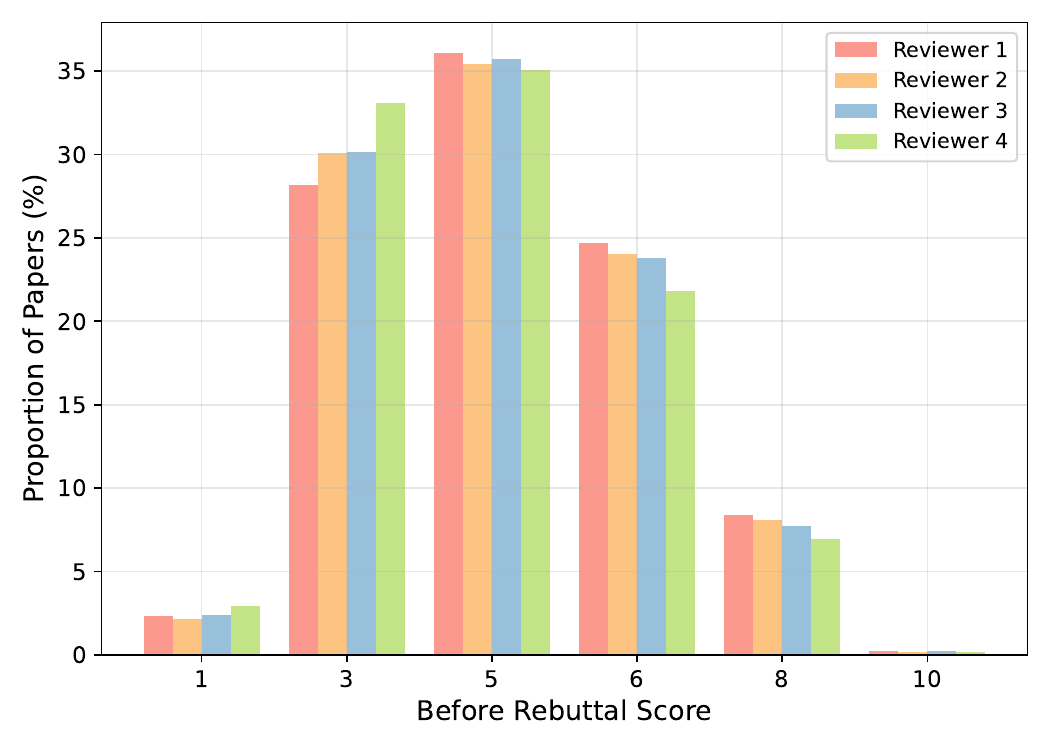}
        \caption{By position (2025)}
        \label{fig:reviewer_time_order_bar_2025_position}
    \end{subfigure}
    \hfill
    \begin{subfigure}[t]{0.23\textwidth}
        \centering
        \includegraphics[width=\textwidth]{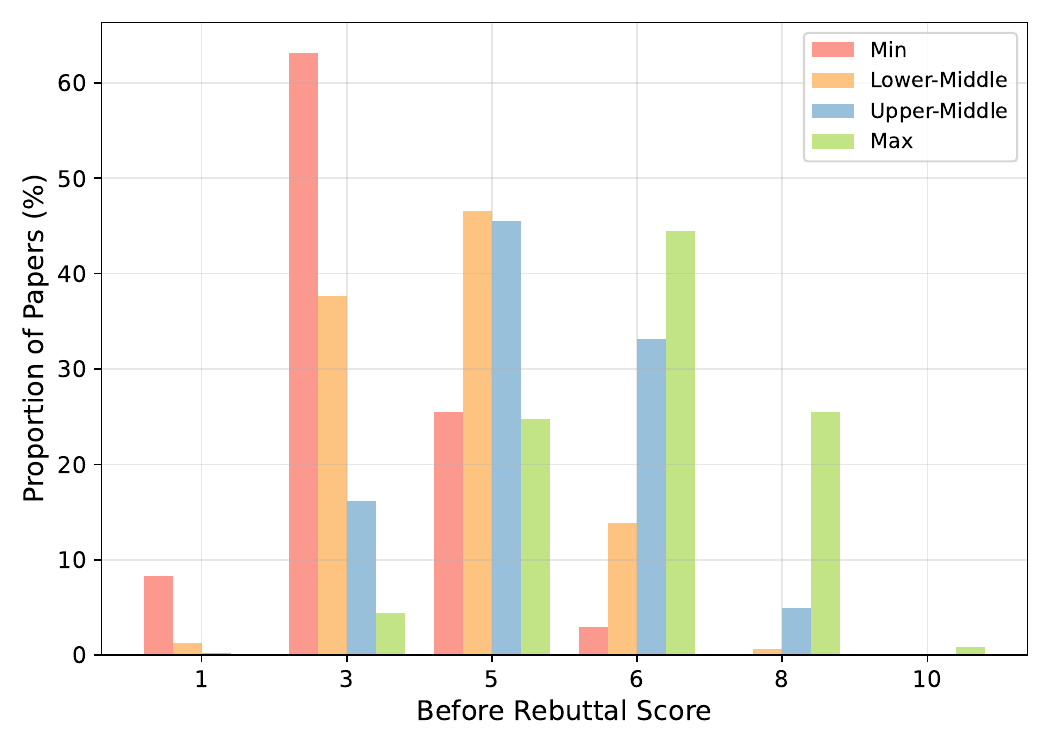}
        \caption{By score rank (2025)}
        \label{fig:reviewer_time_order_bar_2025_minmax}
    \end{subfigure}
    \hfill
    \begin{subfigure}[t]{0.23\textwidth}
        \centering
        \includegraphics[width=\textwidth]{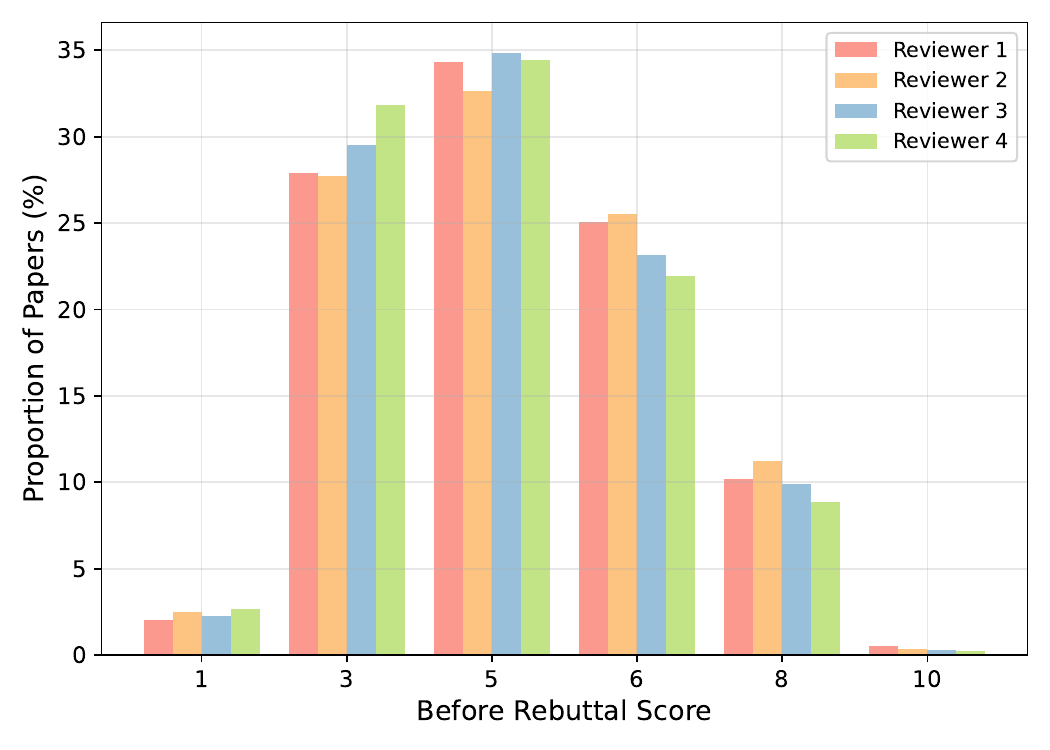}
        \caption{By position (2024)}
        \label{fig:reviewer_time_order_bar_2024_position}
    \end{subfigure}
    \hfill
    \begin{subfigure}[t]{0.23\textwidth}
        \centering
        \includegraphics[width=\textwidth]{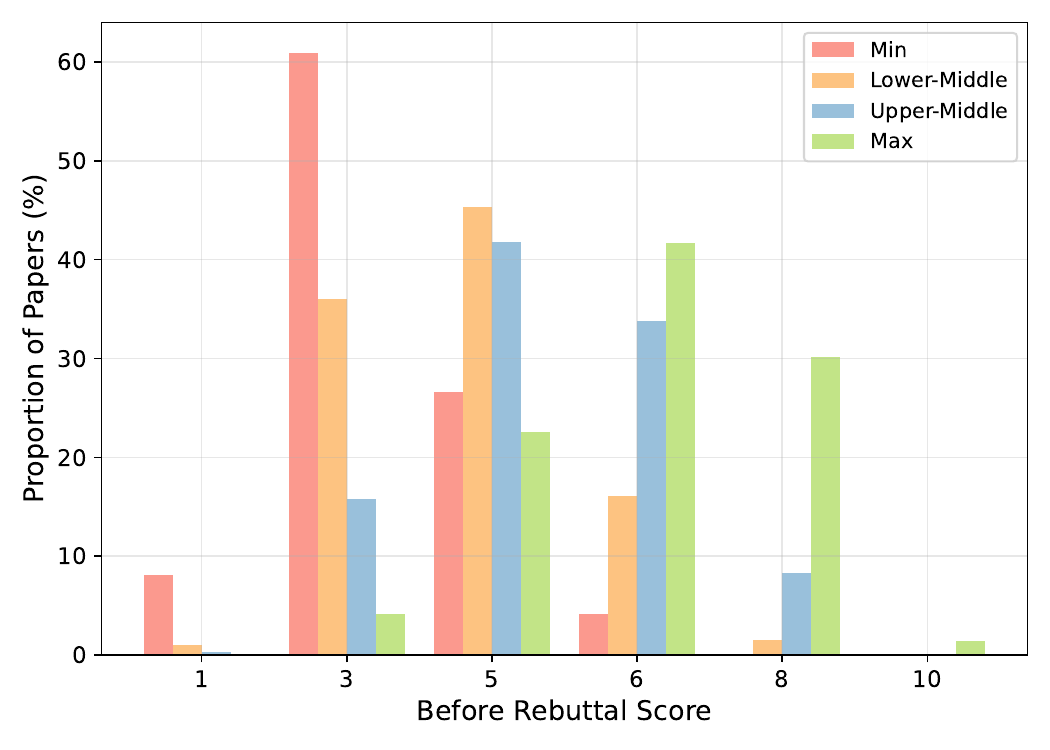}
        \caption{By score rank (2024)}
        \label{fig:reviewer_time_order_bar_2024_minmax}
    \end{subfigure}

    \caption{Distribution of before-rebuttal reviewer scores across papers, grouped either by reviewer position or score rank for ICLR 2025 and ICLR 2024.}
    \label{fig:reviewer_time_order_bar_all}
\end{figure}

%% file: figures/author_reviewer_activity.tex
\begin{figure}[h]
    \centering
    \begin{subfigure}[t]{0.45\textwidth}
        \centering
        \includegraphics[width=\textwidth]{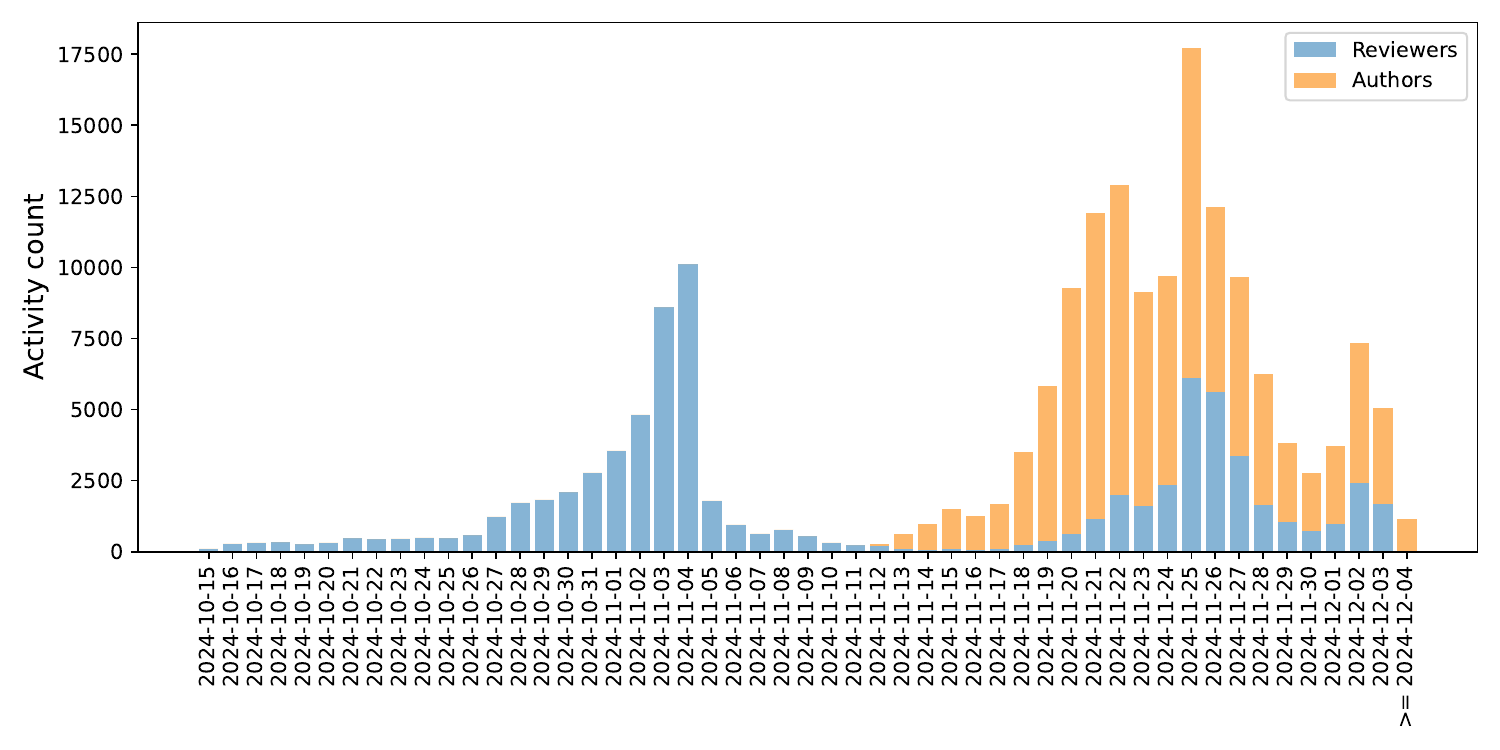}
        \caption{Author and reviewer activity in peer review.}
        \label{fig:author_reviewer_activity_2025}
    \end{subfigure}
    \hfill
    \begin{subfigure}[t]{0.45\textwidth}
        \centering
        \includegraphics[width=\textwidth]{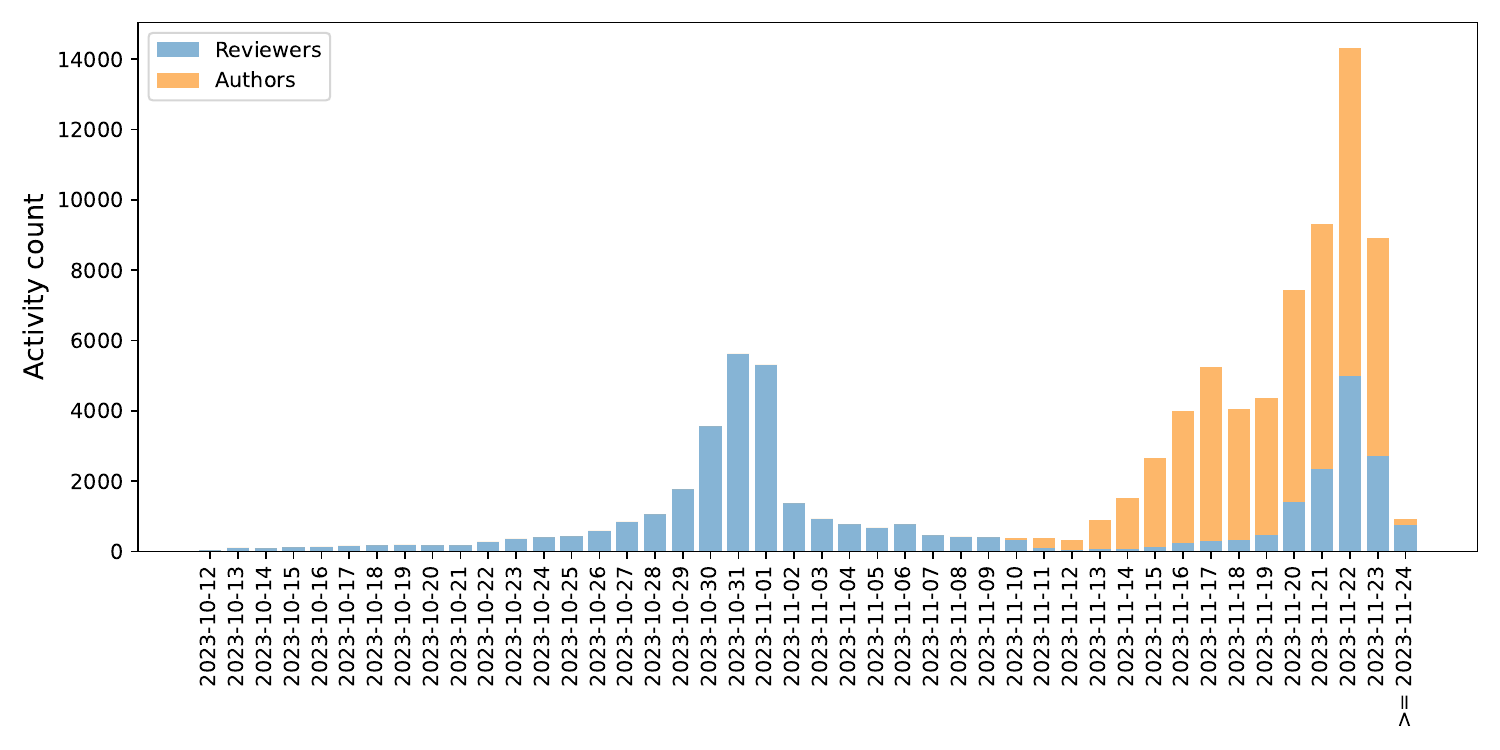}
        \caption{Author and reviewer activity in peer review}
    \label{fig:author_reviewer_activity_2024}
    \end{subfigure}
    
    \caption{Author and reviewer activity in ICLR peer review (2025, left; 2024, right).}
    \label{fig:author_reviewer_activity}
\end{figure}

%% file: figures/author_activity_first_message.tex
\begin{figure}[ht]
    \centering

    \begin{subfigure}[t]{0.45\textwidth}
        \centering
        \includegraphics[width=\textwidth]{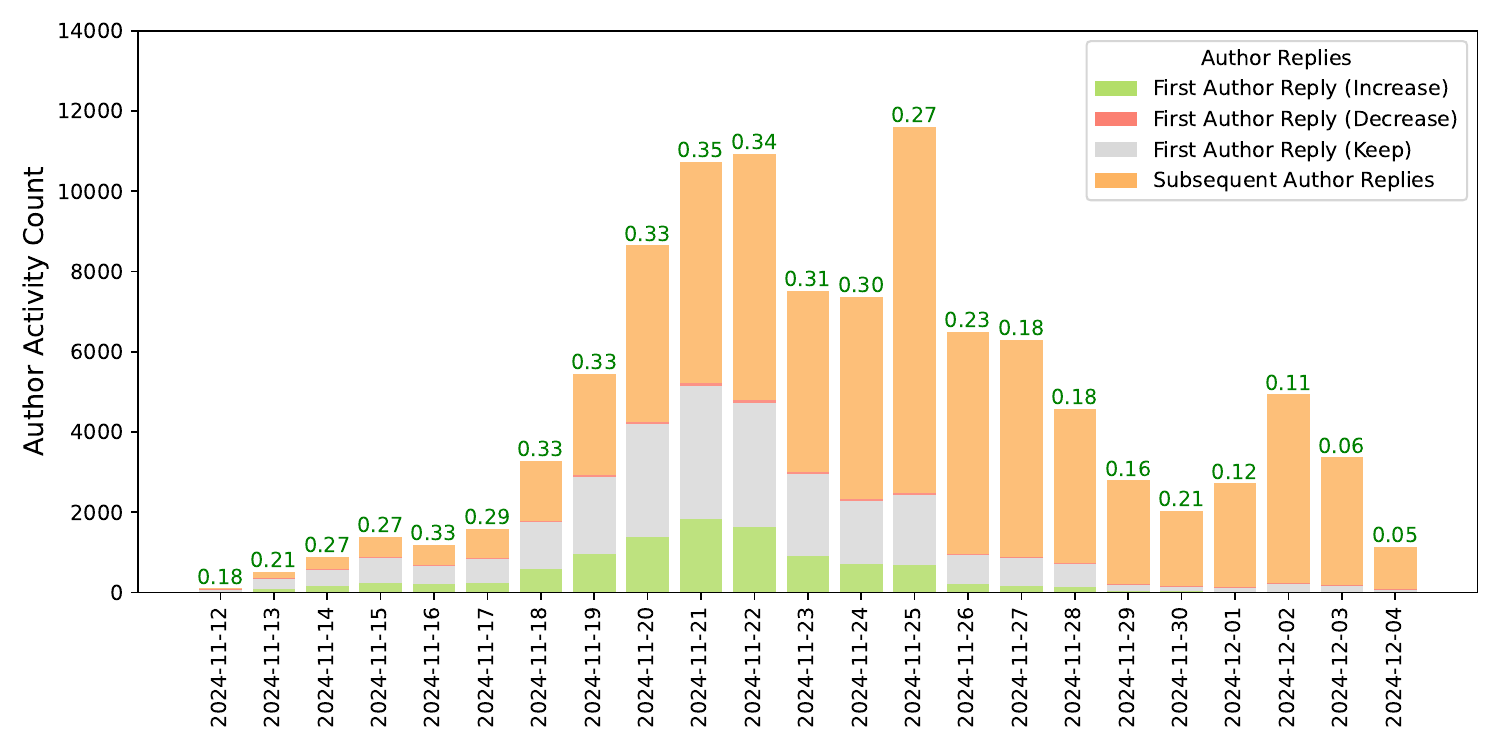}
        \caption{Author activity in rebuttal.}
        \label{fig:author_activity_first_message_2025}
    \end{subfigure}
    \hfill
    \begin{subfigure}[t]{0.45\textwidth}
        \centering
        \includegraphics[width=\textwidth]{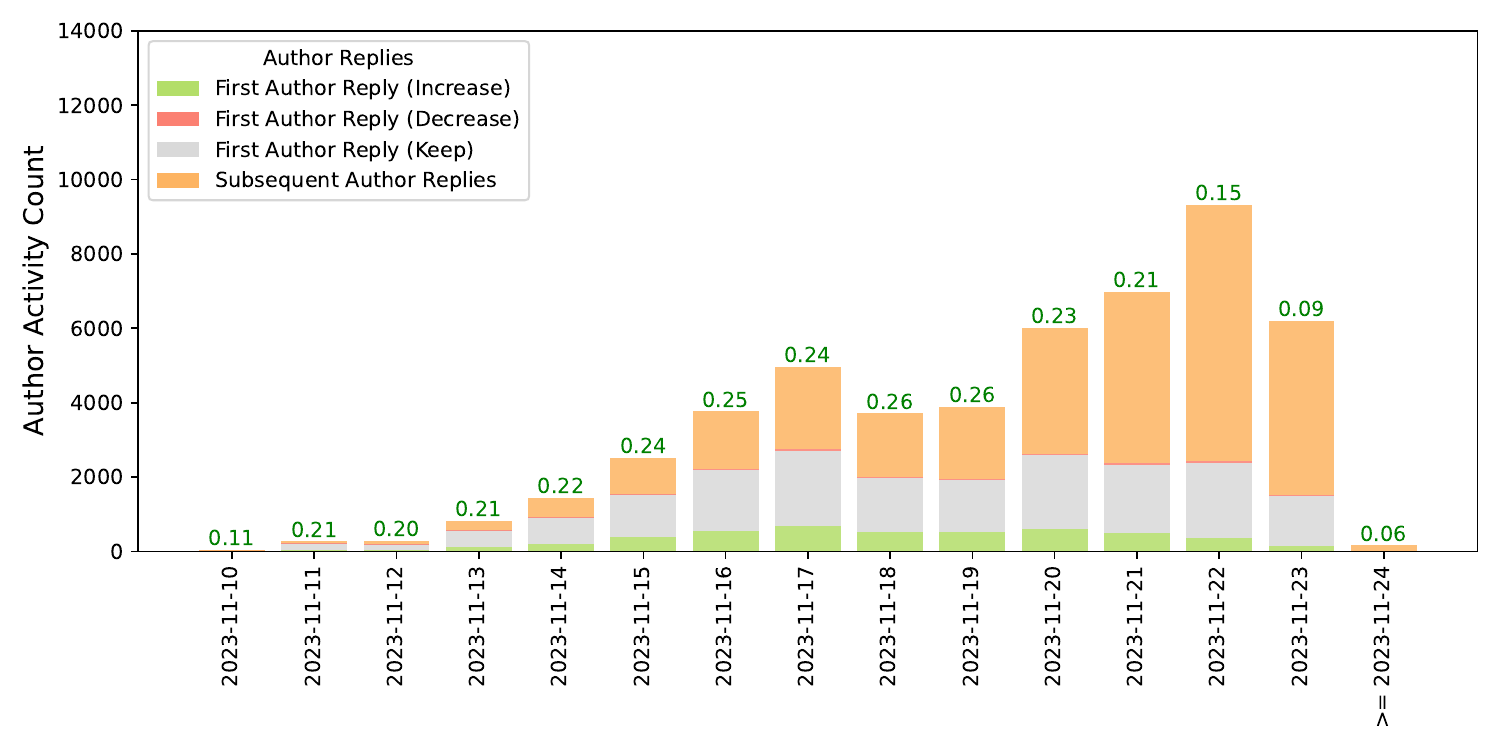}
        \caption{Author activity in rebuttal}
    \label{fig:author_activity_first_message_2024}
    \end{subfigure}
    \caption{Author activity in rebuttal (2025, left; 2024, right). The figure show author first messages (colored by whether reviewers later changed their scores) versus other messages. Numbers above bars indicate the percentage of first messages leading to score increases later.}
    \label{fig:author_activity_first_message}
\end{figure}

%% file: figures/last_modification_date.tex
\begin{figure}[ht]
    \centering
    \begin{subfigure}[t]{0.45\textwidth}
        \centering
        \includegraphics[width=\textwidth]{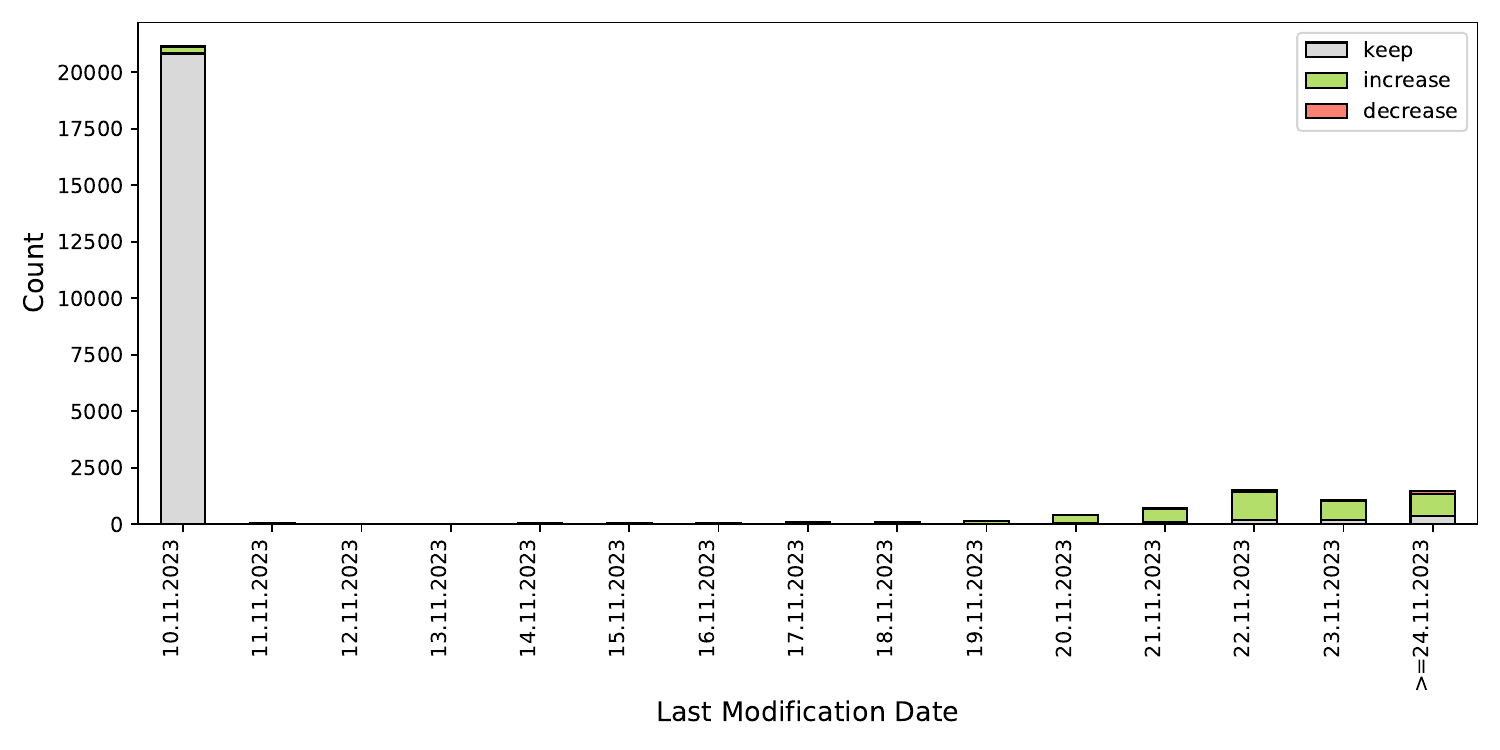}
        \caption{ICLR 2024}
        \label{fig:last_modification_date_2024}
    \end{subfigure}
    \hfill
    \begin{subfigure}[t]{0.45\textwidth}
        \centering
        \includegraphics[width=\textwidth]{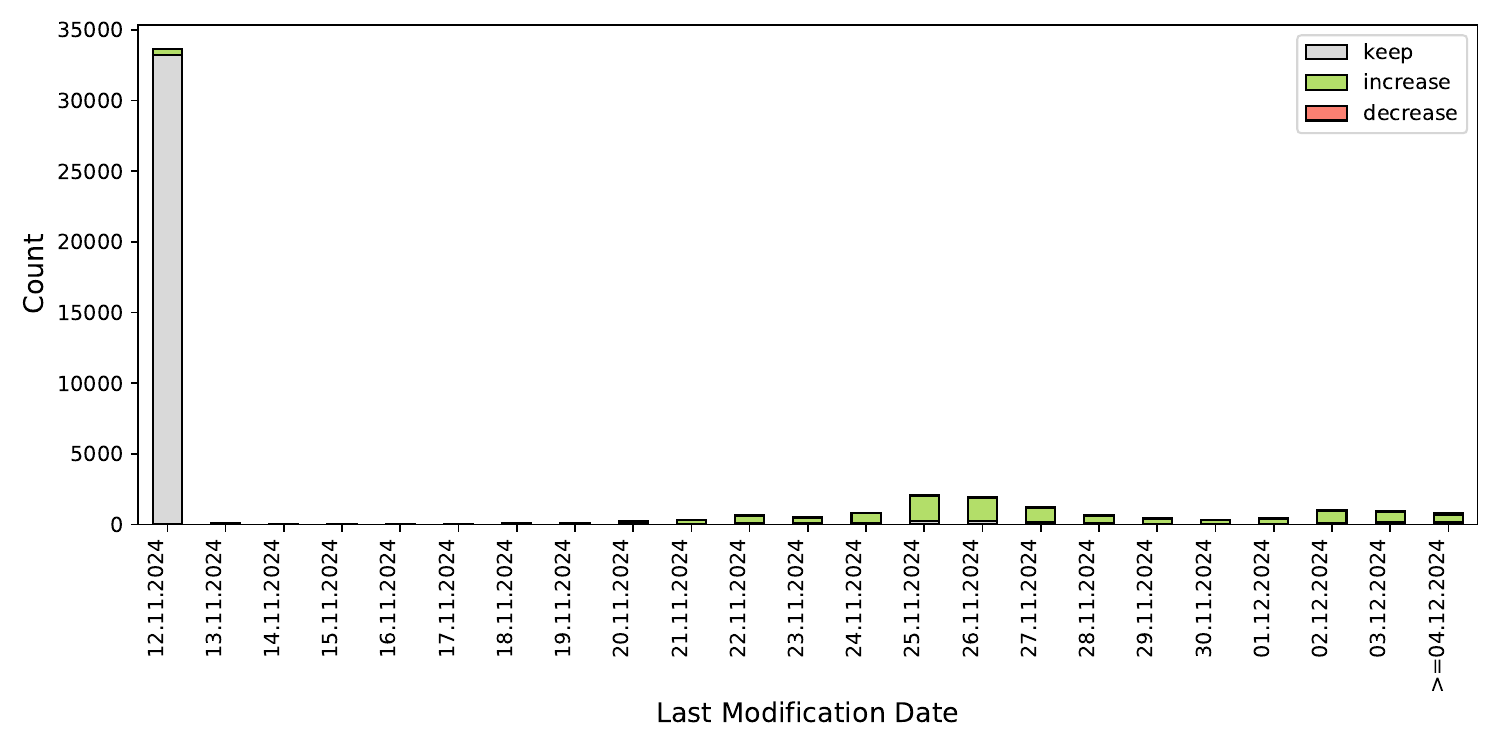}
        \caption{ICLR 2025}
    \label{fig:last_modification_date_2025}
    \end{subfigure}
    \caption{Last modification date of reviews, with the first day corresponding to the release day of the reviews. The colors indicate the review score changes.}
    \label{fig:last_modification_date}
\end{figure}

%% file: tables/reviewer_venue_distance.tex
\begin{table}[ht]
\centering
\caption{Average reviewer disagreement before and after rebuttals, by submission category and year. The table reports absolute ($\Delta$) and relative (Rel. $\Delta\%$) changes in disagreement, with lower values after rebuttals indicating higher peer pressure.
}
\label{tab:reviewer_venue_distance}
\footnotesize
\renewcommand{\arraystretch}{1.2}
\vspace{4pt}
\resizebox{0.9\textwidth}{!}{ 
\begin{tabular}{lc|SSSSSS|S}
\toprule
\textbf{Year} & \textbf{Phase} &
\multicolumn{7}{c}{\textbf{Submission Category}} \\
\cmidrule(lr){3-9} 
 & &
 {Spotlight} & {Oral} & {Poster} & {Rejected} &
 {Withdrawn} & {Desk Rejected} & {\textbf{Total}} \\
\hline
\multirow{2}{*}{\textbf{2025}} 
 & Before & 0.1888 & 0.1715 & 0.1645 & 0.1677 & 0.1502 & 0.0199 & 0.1623 \\
 & After  & 0.1398 {\small ↓} & 0.0889 {\small ↓} & 0.1387 {\small ↓} & 
             0.1579 {\small ↓} & 0.1489 {\small ↓} & 0.0166 {\small ↓} & 0.1478 {\small ↓} \\
 & $\Delta$ & \text{-0.0490} & \text{-0.0826} & \text{-0.0258} & \text{-0.0098} & {-0.0013} & {-0.0033} & \text{-0.0145} \\
 & Rel. $\Delta$\% & \text{-25.96} & \text{-48.16} & \text{-15.68} & \text{-5.86} & \text{-0.88} & \text{-16.49} & \text{-8.92}  \\
\hline
\multirow{2}{*}{\textbf{2024}} 
 & Before & 0.1899 & 0.1690 & 0.1660 & 0.1651 & 0.1480 & 0.0636 & 0.1620 \\
 & After  & 0.1357 {\small ↓} & 0.0992 {\small ↓} & 0.1383 {\small ↓} & 
             0.1535 {\small ↓} & 0.1449 {\small ↓} & 0.0522 {\small ↓} & 0.1456 {\small ↓} \\
 & $\Delta$ & \text{-0.0542} & \text{-0.0699} & \text{-0.0277} & \text{-0.0116} & {-0.0031} & {-0.0115} & \text{-0.0164} \\
 & Rel. $\Delta$\% & \text{-28.54} & \text{-41.33} & \text{-16.69} & \text{-7.00} & \text{-2.09} & \text{-18.00} & \text{-10.12} \\
\bottomrule
\end{tabular}
}
\end{table}

%% file: figures/fig_review_scores_and_heatmap.tex
\begin{figure}[t]
\centering
\begin{minipage}{0.48\textwidth} % left: table
\begin{table}[H] % keeps it as a table
\centering
\footnotesize
\caption{Top feature importances. Weakness features are marked with $\times$, strengths with $\checkmark$. The results are the mean\textsubscript{\text{(std)}} over 10 independent runs.}
\label{tab:feature_importance}
\resizebox{\linewidth}{!}{%
\begin{tabular}{lcc}
\toprule
\textbf{Type} & \textbf{Feature} & \textbf{Avg. $|$Coef.$|$} \\
\midrule 
$\times$ & Novelty \& Contribution & 0.47\textsubscript{(0.01)} \\
$\times$ & Experiments \& Evaluation & 0.32\textsubscript{(0.01)} \\
$\times$ & Word Count & 0.25\textsubscript{(0.01)} \\
$\checkmark$ & Writing \& Presentation & 0.23\textsubscript{(0.01)} \\
$\times$ & Methodology \& Technical Soundness & 0.23\textsubscript{(0.01)} \\
$\times$ & Motivation & 0.20\textsubscript{(0.01)} \\
$\checkmark$ & Methodology \& Technical Soundness & 0.20\textsubscript{(0.01)} \\
$\checkmark$ & Word Count & 0.19\textsubscript{(0.01)} \\
$\checkmark$ & Results & 0.19\textsubscript{(0.01)} \\
$\checkmark$ & Experiments \& Evaluation & 0.18\textsubscript{(0.01)} \\
$\checkmark$ & Novelty \& Contribution & 0.18\textsubscript{(0.01)} \\
\bottomrule
\end{tabular}%
}
\end{table}
\end{minipage}%
\hfill
\begin{minipage}{0.45\textwidth} % right: figure
\centering
\includegraphics[width=\linewidth]{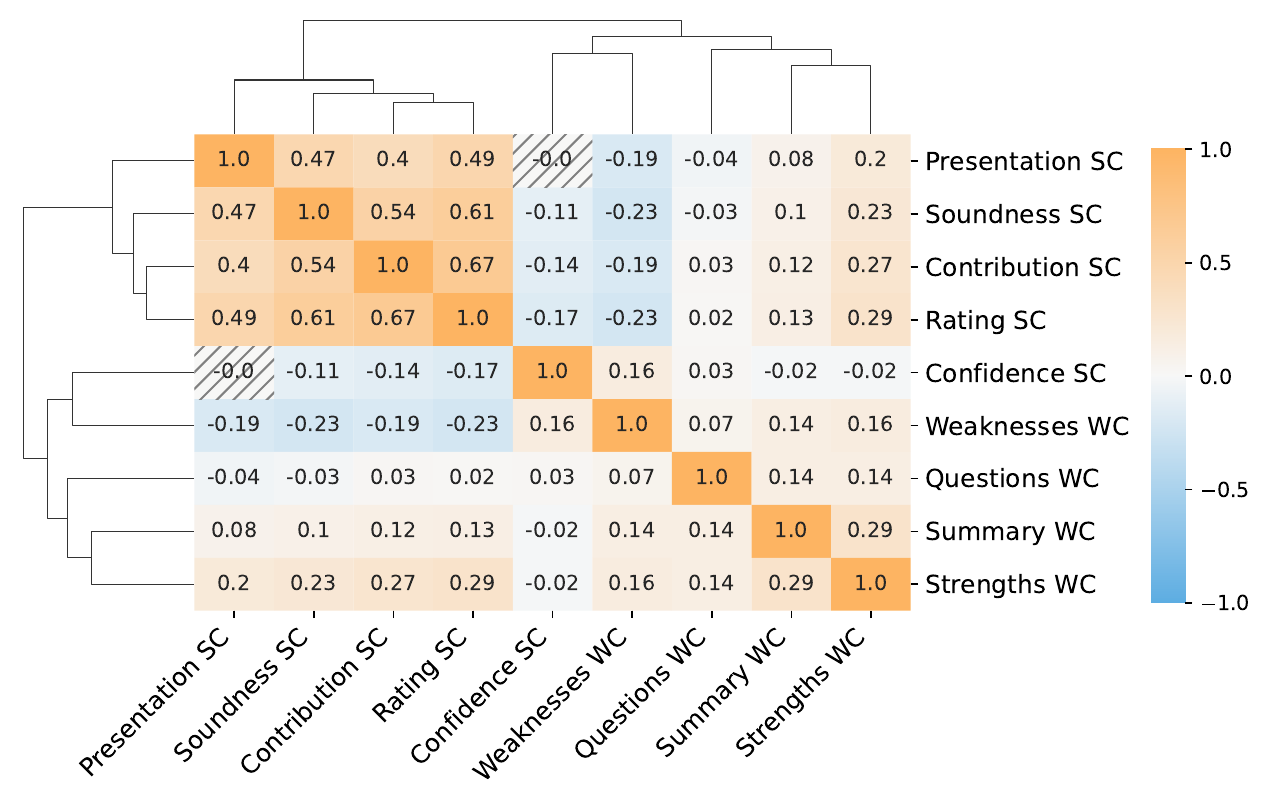} % replace with your figure file
\caption{Correlation heatmap of after rebuttal score (SC) and word count (WC). Non-hatched cells indicate $p<0.001$.}
\label{fig:correlation_heatmap}
\end{minipage}
\end{figure}

%% file: tables/stats_per_category.tex
\begin{table}[t]
\centering
\footnotesize
\caption{Average reviewer scores and word counts by submission category. The abbreviations are: Sound. (Soundness), Present. (Presentation), Contrib. (Contribution), Confid. (Confidence), and Summ. (Summary). Bold numbers indicate the highest value in each column.\vspace{5pt}}
\label{tab:stats_per_category}
 \resizebox{0.99\textwidth}{!}{%
\begin{tabular}{lrrrrrrrrr}
\toprule
\textbf{Category} & \multicolumn{5}{c}{\textbf{Scores}} & \multicolumn{4}{c}{\textbf{Word Counts}} \\
\cmidrule(lr){2-6} \cmidrule(lr){7-10}
 & Sound. & Present. & Contrib. & Rating & Confid. & Summ. & Strengths & Weaknesses & Questions \\
\midrule
                    Oral &          \textbf{3.21} &             \textbf{3.19} &             \textbf{3.13} &       \textbf{7.74} &           3.58 &      \textbf{103.86} &         \textbf{90.32} &         167.50 &        \textbf{104.67} \\
               Spotlight &          3.09 &             3.07 &             2.95 &       7.24 &           3.56 &       98.74 &         80.54 &         160.91 &         92.59 \\
Poster &          2.87 &             2.87 &             2.67 &       6.26 &           3.57 &       92.80 &         72.15 &         174.53 &         88.03 \\
Desk Rejected &          2.57 &             2.58 &             2.40 &       5.14 &           3.72 &       80.27 &         52.02 &         182.75 &         78.28 \\
Rejected &          2.51 &             2.55 &             2.27 &       4.79 &           3.64 &       86.62 &         60.39 &         205.63 &         88.36 \\
Withdrawn &          2.34 &             2.40 &             2.08 &       4.07 &           \textbf{3.81} &       82.00 &         55.73 &         \textbf{220.34} &         82.83 \\
\bottomrule
\end{tabular}
}\vspace{10pt}
\end{table}

%% file: tables/score_changes_summary.tex
\begin{table}[t]
\centering
\footnotesize
\caption{Score changes in ICLR 2025 reviews across different score fields based on the type of overall score change. `Any'' indicates any change, while `Any Except Rating'' excludes changes to the overall rating.}
\label{tab:score_changes_summary}
\resizebox{0.99\textwidth}{!}{%
\footnotesize
\renewcommand{\arraystretch}{1.2}
\begin{tabular}{l|c|cccccccc}
\toprule
\textbf{Type} & \textbf{\# No Score Change} & \multicolumn{7}{c}{\textbf{\# Score Change}} \\
\cmidrule(lr){3-9}
 &  & Any & Rating & Confidence & Soundness & Contribution & Presentation & Any Except Rating \\
\hline
{Decrease}   & 0      & 640    & 640    & 116        & 114       & 116         & 77           & 267  \\
{Increase}   & 0      & 10728  & 10728  & 717        & 1325      & 1303        & 935          & 3021 \\
{Keep} & 33913  & 1072   & 0      & 579        & 265       & 214         & 281          & 1072 \\
\bottomrule
\end{tabular}%
}
\end{table}

%% file: tables/score_change_prediction.tex
\begin{table}[t]
\centering
\caption{F1-scores per class and macro F1 for 2- and 3-class classification. The results are the mean\textsubscript{\text{(std)}} over 10 independent runs.}
\vspace{3pt}
\label{tab:score_change_prediction}
\resizebox{0.75\textwidth}{!}{%
\begin{tabular}{l l c c c}
\toprule
\textbf{Setting} & \textbf{Class} & \textbf{F1-score} & \textbf{Macro F1} & \textbf{Random Macro F1} \\
\midrule
\multirow{2}{*}{2-class (Decrease/Keep vs Increase)} & Decrease/Keep & 0.80\textsubscript{(0.01)} & \multirow{2}{*}{0.69\textsubscript{(0.02)}} & \multirow{2}{*}{0.50} \\
                                           & Increase      & 0.57\textsubscript{(0.03)} &        &                          \\
\midrule
\multirow{3}{*}{3-class (Decrease, Keep, Increase)}  & Decrease      & 0.18\textsubscript{(0.05)} & \multirow{3}{*}{0.51\textsubscript{(0.02)}} & \multirow{3}{*}{0.33} \\
                                           & Increase      & 0.59\textsubscript{(0.02)} &        &                          \\
                                           & Keep     & 0.76\textsubscript{(0.02)} &        &                          \\
\bottomrule
\end{tabular}%
}
\end{table}

%% file: tables/rebuttal_strategy.tex
\begin{table}[t]
\centering
\footnotesize
\caption{Top features and coefficients for DEC, INC, and KEEP classes. For All it shows the mean absolute coefficient across classes. The results are the mean\textsubscript{\text{(std)}} of 10 independent runs.}\vspace{3pt}
\label{tab:top_features_coef}
\resizebox{0.99\textwidth}{!}{%
\begin{tabular}{lcccc}
\toprule
Feature & All (Avg. $|$Coef.$|$) & DEC & INC & KEEP \\
\midrule
Overall rating score (before the rebuttal) & 0.67\textsubscript{(0.03)} & \cellcolor{tabgreen!91}0.92\textsubscript{(0.04)} & \cellcolor{tabred!100}-1.01\textsubscript{(0.03)} & \cellcolor{tabgreen!7}0.08\textsubscript{(0.02)} \\
Mean overall rating score of other reviewers (before the rebuttal)  & 0.37\textsubscript{(0.02)} & \cellcolor{tabred!55}-0.56\textsubscript{(0.03)} & \cellcolor{tabgreen!54}0.55\textsubscript{(0.01)} & \cellcolor{tabgreen!0}0.01\textsubscript{(0.02)} \\
Reviewer engagement (number of notes) & 0.25\textsubscript{(0.02)} & \cellcolor{tabgreen!16}0.17\textsubscript{(0.03)} & \cellcolor{tabgreen!19}0.20\textsubscript{(0.02)} & \cellcolor{tabred!36}-0.37\textsubscript{(0.02)} \\
Generic/vague defense (strategy) & 0.19\textsubscript{(0.15)} & \cellcolor{tabgreen!2}0.03\textsubscript{(0.17)} & \cellcolor{tabred!28}-0.29\textsubscript{(0.15)} & \cellcolor{tabgreen!25}0.26\textsubscript{(0.12)} \\
Contribution score (before the rebuttal) & 0.19\textsubscript{(0.02)} & \cellcolor{tabred!26}-0.27\textsubscript{(0.02)} & \cellcolor{tabgreen!27}0.28\textsubscript{(0.01)} & \cellcolor{tabred!0}-0.01\textsubscript{(0.01)} \\
Evidence backed clarification (strategy) & 0.15\textsubscript{(0.10)} & \cellcolor{tabred!2}-0.03\textsubscript{(0.12)} & \cellcolor{tabgreen!22}0.23\textsubscript{(0.10)} & \cellcolor{tabred!19}-0.20\textsubscript{(0.09)} \\
Bare agreement/disagreement (strategy subcategory) & 0.13\textsubscript{(0.09)} & \cellcolor{tabred!4}-0.05\textsubscript{(0.10)} & \cellcolor{tabgreen!18}0.19\textsubscript{(0.09)} & \cellcolor{tabred!13}-0.14\textsubscript{(0.07)} \\
Evasion (stance subcategory) & 0.10\textsubscript{(0.04)} & \cellcolor{tabred!5}-0.06\textsubscript{(0.03)} & \cellcolor{tabred!9}-0.10\textsubscript{(0.04)} & \cellcolor{tabgreen!14}0.15\textsubscript{(0.07)} \\
Disagree (stance) & 0.10\textsubscript{(0.12)} & \cellcolor{tabgreen!1}0.02\textsubscript{(0.05)} & \cellcolor{tabred!14}-0.15\textsubscript{(0.13)} & \cellcolor{tabgreen!12}0.13\textsubscript{(0.18)} \\
Method details (strategy subcategory) & 0.10\textsubscript{(0.06)} & \cellcolor{tabgreen!0}0.01\textsubscript{(0.07)} & \cellcolor{tabred!14}-0.15\textsubscript{(0.06)} & \cellcolor{tabgreen!13}0.14\textsubscript{(0.05)} \\
Future promise (strategy subcategory) & 0.10\textsubscript{(0.08)} & \cellcolor{tabgreen!4}0.05\textsubscript{(0.09)} & \cellcolor{tabgreen!8}0.09\textsubscript{(0.08)} & \cellcolor{tabred!14}-0.15\textsubscript{(0.07)} \\
Soundness score (before the rebuttal) & 0.10\textsubscript{(0.02)} & \cellcolor{tabred!11}-0.12\textsubscript{(0.03)} & \cellcolor{tabgreen!13}0.14\textsubscript{(0.02)} & \cellcolor{tabred!2}-0.03\textsubscript{(0.02)} \\
Broad assertion (strategy subcategory) & 0.09\textsubscript{(0.08)} & \cellcolor{tabgreen!3}0.04\textsubscript{(0.10)} & \cellcolor{tabgreen!8}0.09\textsubscript{(0.08)} & \cellcolor{tabred!13}-0.14\textsubscript{(0.07)} \\
Reviewer confidence score (before the rebuttal) & 0.09\textsubscript{(0.02)} & \cellcolor{tabgreen!11}0.12\textsubscript{(0.02)} & \cellcolor{tabred!12}-0.13\textsubscript{(0.01)} & \cellcolor{tabgreen!0}0.01\textsubscript{(0.01)} \\
Evasion (strategy subcategory) & 0.08\textsubscript{(0.03)} & \cellcolor{tabgreen!11}0.12\textsubscript{(0.04)} & \cellcolor{tabred!5}-0.06\textsubscript{(0.03)} & \cellcolor{tabred!6}-0.07\textsubscript{(0.03)} \\
\bottomrule
\end{tabular}%
}
\end{table}

%% file: figures/taxonomy_weakness.tex
\begin{figure}[htbp]
    \centering
\resizebox{0.85\textwidth}{!}{% scale to page width
\begin{forest}
for tree={
    grow'=0,                   % horizontal tree
    draw,
    rounded corners,
    node options={align=center},
    font=\scriptsize,           % smaller font
    edge={thick},
    parent anchor=east,
    child anchor=west,
    anchor=west,
    l sep=5mm,                  % smaller level separation
    s sep=1mm                   % smaller sibling separation
}
[Rating Score (Before Rebuttal)
    [Rating 1, fill=red!20
        [Writing \& Presentation (2.18)
            [Too much jargon / unclear wording (0.46), fill=red!10]
            [Unclear problem definition (0.41), fill=red!15]
            [Poor structure / flow (0.35), fill=red!20]
        ]
        [Experiments \& Evaluation (1.87)
            [Insufficient or weak baselines (0.49), fill=orange!10]
            [Reproducibility issues (0.28), fill=orange!15]
            [Too few datasets / limited domain (0.24), fill=orange!20]
        ]
        [Methodology \& Technical Soundness (1.70)
            [Unclear algorithmic description (0.55), fill=yellow!10]
            [Incorrect/unrealistic assumptions (0.48), fill=yellow!15]
            [Weak theoretical justification (0.40), fill=yellow!20]
        ]
        [Novelty \& Contribution (1.01)
            [Overlapping with prior work (0.30), fill=green!10]
            [Lack of originality (0.28), fill=green!15]
            [Lack of clear contribution (0.26), fill=green!20]
        ]
    ]
    [Rating 3, fill=blue!20
        [Experiments \& Evaluation (1.91)
            [Insufficient or weak baselines (0.49), fill=orange!10]
            [Too few datasets / limited domain (0.27), fill=orange!15]
            [Missing ablation tests (0.21), fill=orange!20]
        ]
        [Methodology \& Technical Soundness (1.50)
            [Unclear algorithmic description (0.41), fill=yellow!10]
            [Weak theoretical justification (0.38), fill=yellow!15]
            [Incorrect/unrealistic assumptions (0.38), fill=yellow!20]
        ]
        [Writing \& Presentation (1.32)
            [Too much jargon / unclear wording (0.26), fill=red!10]
            [Unclear problem definition: (0.21), fill=red!15]
            [Weak/missing figures \& tables (0.21), fill=red!20]
        ]
        [Novelty \& Contribution (1.01)
            [Lack of clear contribution (0.31), fill=green!10]
            [Lack of originality (0.30), fill=green!15]
            [Overlapping with prior work (0.22), fill=green!20]
        ]
    ]
    [Rating 5, fill=purple!20
        [Experiments \& Evaluation (1.77)
            [Insufficient or weak baselines (0.39), fill=orange!10]
            [Too few datasets / limited domain (0.26), fill=orange!15]
            [Missing ablation tests (0.23), fill=orange!20]
        ]
        [Methodology \& Technical Soundness (1.26)
            [Unclear algorithmic description (0.35), fill=yellow!10]
            [Weak theoretical justification (0.29), fill=yellow!15]
            [Incorrect/unrealistic assumptions (0.28), fill=yellow!20]
        ]
        [Writing \& Presentation (1.01)
            [Weak/missing figures \& tables (0.18), fill=red!10]
            [Too much jargon / unclear wording (0.17), fill=red!15]
            [Unclear problem definition (0.15), fill=red!20]
        ]
    ]
    [Rating 6, fill=teal!20
        [Experiments \& Evaluation (1.42)
            [Insufficient or weak baselines (0.27), fill=orange!10]
            [Too few datasets / limited domain (0.23), fill=orange!15]
            [Missing ablation tests (0.20), fill=orange!20]
        ]
        [Methodology \& Technical Soundness (1.06)
            [Unclear algorithmic description (0.30), fill=yellow!10]
            [Incorrect or unrealistic assumptions (0.21), fill=yellow!15]
            [Weak theoretical justification (0.20), fill=yellow!20]
        ]
    ]
    [Rating 8, fill=gray!20
        [Experiments \& Evaluation (1.10)
            [Too few datasets / limited domain (0.18), fill=orange!10]
            [Insufficient or weak baselines (0.18), fill=orange!15]
            [Missing ablation tests (0.15), fill=orange!20]
        ]
    ]
    [Rating 10, fill=pink!20
        [Writing \& Presentation (1.06)
            [Too much jargon / unclear wording (0.25), fill=red!10]
            [Weak/missing figures \& tables (0.21), fill=red!15]
            [Contribution not clearly communicated (0.20), fill=red!20]
        ]
    ]
]
\end{forest}
} % end resizebox
\caption{Taxonomy of weaknesses raised by reviewers at ICLR 2024 and 2025, categorized by rating scores. Values in parentheses represent the number of mentions per review. We report only categories with values above 1.0 and, for each, select the top three subcategories.}
\label{fig:taxonomy_weakness}
\end{figure}

%% file: figures/taxonomy_strength.tex
\begin{figure}[htbp]
    \centering
\resizebox{0.95\textwidth}{!}{% scale to page width
\begin{forest}
for tree={
    grow'=0,                   % horizontal tree
    draw,
    rounded corners,
    node options={align=center},
    font=\scriptsize,           % smaller font
    edge={thick},
    parent anchor=east,
    child anchor=west,
    anchor=west,
    l sep=5mm,                  % smaller level separation
    s sep=1mm                   % smaller sibling separation
}
[Rating Score (Before Rebuttal)
    [Rating 1, fill=red!20
    ]
    [Rating 3, fill=blue!20
    ]
    [Rating 5, fill=purple!20
        [Novelty \& Contribution (0.84)
            [Original idea / high novelty (0.54), fill=green!10]
            [Significant contribution (0.16), fill=green!15]
            [Clear and meaningful improvement over prior work (0.12), fill=green!20]
        ]
    ]
    [Rating 6, fill=teal!20
        [Novelty \& Contribution (0.99)
            [Original idea / high novelty (0.58), fill=green!10]
            [Significant contribution (0.21), fill=green!15]
            [Clear and meaningful improvement over prior work (0.14), fill=green!20]
        ]
        [Methodology \& Technical Soundness (0.91)
            [Elegant and efficient model or approach (0.37), fill=yellow!10]
            [Strong theoretical justification (0.22), fill=yellow!15]
            [Thoughtful and justified design choices (0.17), fill=yellow!20]
        ]
    ]
    [Rating 8, fill=gray!20
        [Novelty \& Contribution (1.17)
            [Original idea / high novelty (0.62), fill=green!10]
            [Significant contribution (0.29), fill=green!15]
            [Clear and meaningful improvement over prior work (0.17), fill=green!20]
        ]
        [Methodology \& Technical Soundness (0.95)
            [Elegant and efficient model or approach (0.35), fill=yellow!10]
            [Strong theoretical justification (0.22), fill=yellow!15]
            [Thoughtful and justified design choices (0.19), fill=yellow!20]
        ]
        [Experiments \& Evaluation (0.90)
            [Realistic experiments / real-world validation (0.28), fill=orange!10]
            [Thorough ablation studies (0.16), fill=orange!15]
            [Large-scale - diverse datasets / broad coverage (0.14), fill=orange!20]
        ]
        [Writing \& Presentation (0.87)
            [Well-structured and organized (0.31), fill=red!10]
            [Clear problem definition (0.14), fill=red!15]
            [Accessible language - minimal jargon (0.14), fill=red!20]
        ]
    ]
    [Rating 10, fill=pink!20
        [Novelty \& Contribution (1.31)
            [Original idea / high novelty (0.66), fill=green!10]
            [Significant contribution (0.37), fill=green!15]
            [Clear and meaningful improvement over prior work (0.18), fill=green!20]
        ]
        [Methodology \& Technical Soundness (1.10)
            [Elegant and efficient model or approach (0.50), fill=yellow!10]
            [Strong theoretical justification (0.24), fill=yellow!15]
            [Thoughtful and justified design choices (0.13), fill=yellow!20]
        ]
        [Results (1.04)
            [Significant or substantial gains (0.49), fill=orange!10]
            [Consistent and explainable findings (0.22), fill=orange!15]
            [Clear and correct interpretation of results (0.18), fill=orange!20]
        ]
        [Writing \& Presentation (0.97)
            [Well-structured and organized (0.32), fill=red!10]
            [Strong and informative figures \& tables (0.18), fill=red!15]
            [Clear problem definition (0.16), fill=red!20]
        ]
        [Experiments \& Evaluation (0.93)
            [Realistic experiments / real-world validation (0.22), fill=orange!10]
            [Large-scale - diverse datasets / broad coverage (0.19), fill=orange!15]
            [Thorough ablation studies (0.12), fill=orange!20]
        ]
    ]
]
\end{forest}
}
\caption{Taxonomy of strengths raised by reviewers at ICLR 2024 and 2025, categorized by rating scores. Values in parentheses represent the number of mentions per review. We report only categories with values above 0.8 and, for each, select the top three subcategories.}
\label{fig:taxonomy_strength}
\end{figure}

%% file: figures/weakness_prompt.tex
\begin{figure*}[t]
\begin{tcolorbox}[wbox]
\textbf{Task Description:} You are an annotator. Your task is to analyze the weaknesses in a computer science paper review. 

For each paragraph in the text, produce a list of annotations. For each annotation: 
\begin{enumerate}[leftmargin=1.5em, itemsep=0pt]
\item Assign one of the main categories from the list below. You must not use ``Other'' unless it is absolutely impossible to map the weakness to any category. Even a very weak or partial connection (as little as 1\%) is enough to assign an existing category.
\item Assign a subcategory: use one of the subcategories defined under the chosen main category. You must always choose a defined subcategory if there is any possible relation at all, no matter how small. Only if it is truly impossible to relate the weakness to any defined subcategory, then use ``Other''.
\item Extract the exact text span that expresses the weakness.
\item If multiple weaknesses exist in one paragraph, annotate each separately in annotation list.
\end{enumerate}
Always prioritize using the predefined categories and subcategories over ``Other''.

\medskip
\textbf{Main categories and their subcategories:}

\begin{enumerate}[leftmargin=1.5em, itemsep=0pt]
  \item \textbf{Novelty \& Contribution}: 
  Lack of originality; Incremental improvement; Lack of clear contribution; Overclaiming novelty; 
  Overlapping with prior work; Work not mature enough for publication; Other.

  \item \textbf{Motivation}: 
  Weak or missing motivation; Problem not well justified as important; 
  No clear real-world or theoretical relevance; Other.

  \item \textbf{Methodology \& Technical Soundness}: 
  Weak theoretical justification; Incorrect or unrealistic assumptions; Overly complicated model; 
  Cherry-picked design choices; Unclear algorithmic description; Scalability or computational impracticality; 
  Lack of consideration of established improvement techniques; Other.

  \item \textbf{Experiments \& Evaluation}: 
  Insufficient or weak baselines; Too few datasets / limited domain or language coverage; 
  Small-scale experiments / lack of real-world validation; Poor generalizability across settings; 
  Missing error or failure analysis; Missing ablation tests; Lack of statistical significance tests; 
  No human evaluation when needed; Evaluation metrics not well justified; 
  Unfair comparisons (own method tuned, others not); Reproducibility issues (missing details, code not available); Other.

  \item \textbf{Results}: 
  Marginal gains / not significant; Overinterpretation of results; Contradictory or unexplained findings; 
  Weak interpretability / lack of explanation; No qualitative examples or case studies; Other.

  \item \textbf{Data}: 
  Dataset not publicly available; Poor data quality / noise not addressed; 
  Missing dataset documentation or statistics; Missing inter-annotator agreement scores; 
  Synthetic or toy datasets only; Other.

  \item \textbf{Writing \& Presentation}: 
  Unclear problem definition; Poor structure / flow; Too much jargon / unclear wording; 
  Weak or missing figures \& tables; Inconsistent terminology; Grammar, typos, formatting issues; 
  Contribution not clearly communicated; Other.

  \item \textbf{Broader Impact, Ethics \& Relevance}: 
  No discussion of societal risks or ethical implications; Unrealistic claims of impact; 
  Ignoring potential biases, fairness, or safety issues; Other.

  \item \textbf{References \& Related Work}: 
  Missing related work; Missing important prior work; Outdated references; Other.

  \item \textbf{Fit \& Scope for Venue}: 
  Mismatch with venue; Other.

  \item \textbf{Other}: 
  Other.
\end{enumerate}

\medskip
\textbf{Output Format:}  
Return the annotations as a JSON list, where each element corresponds to a paragraph.  
Put the JSON list in the \texttt{<START\_LIST>} \texttt{<END\_LIST>}.

{\scriptsize
\begin{verbatim}
{
 "paragraph": "<original paragraph>",
 "annotation_list": [
   {
     "main_category": "<one of the 11 main categories>",
     "subcategory": "<one of the subcategories for the selected main category>",
     "source": "<exact phrase/sentence/paragraph that expresses the weakness>"
   }
 ]
}
\end{verbatim}
}
\end{tcolorbox}
\caption{Weakness annotation prompt}\label{fig:weakness_prompt}
\end{figure*}

%% file: figures/strength_prompt.tex
\begin{figure*}[t]
\begin{tcolorbox}[gbox]
\textbf{Task Description:} You are an annotator. Your task is to analyze the strengths in a computer science paper review. 

For each paragraph in the text, produce a list of annotations. For each annotation: 
\begin{enumerate}[leftmargin=1.5em, itemsep=0pt]
\item Assign one of the main categories from the list below. You must not use ``Other'' unless it is absolutely impossible to map the strength to any category. Even a very weak or partial connection (as little as 1\%) is enough to assign an existing category.
\item Assign a subcategory: use one of the subcategories defined under the chosen main category. You must always choose a defined subcategory if there is any possible relation at all, no matter how small. Only if it is truly impossible to relate the strength to any defined subcategory, then use ``Other''.
\item Extract the exact text span that expresses the strength.
\item If multiple strengths exist in one paragraph, annotate each separately in annotation list.
\end{enumerate}
Always prioritize using the predefined categories and subcategories over ``Other''.

\medskip
\textbf{Main categories and their subcategories:}

\begin{enumerate}[leftmargin=1.5em, itemsep=0pt]
  \item \textbf{Novelty \& Contribution}: 
  Original idea / high novelty; Significant contribution; Clear and meaningful improvement over prior work; 
  Strong theoretical or practical impact; Well-scoped and mature work; Other.

  \item \textbf{Motivation}: 
  Well-motivated problem; Clearly important problem; Strong real-world or theoretical relevance; Other.

  \item \textbf{Methodology \& Technical Soundness}: 
  Strong theoretical justification; Realistic and valid assumptions; Elegant and efficient model or approach; 
  Thoughtful and justified design choices; Clear and reproducible algorithmic description; 
  Scalable and computationally practical; Builds on established techniques effectively; Other.

  \item \textbf{Experiments \& Evaluation}: 
  Strong baselines; Large-scale, diverse datasets / broad coverage; Realistic experiments / real-world validation; 
  Good generalizability across settings; Detailed error or failure analysis; Thorough ablation studies; 
  Statistical significance tested; Human evaluation included when needed; Well-justified evaluation metrics; 
  Fair comparisons; Reproducible (code and details provided); Other.

  \item \textbf{Results}: 
  Significant or substantial gains; Clear and correct interpretation of results; Consistent and explainable findings; 
  Strong interpretability / well-explained results; Includes qualitative examples or case studies; Other.

  \item \textbf{Data}: 
  High-quality datasets; Publicly available data; Detailed dataset documentation and statistics; 
  Inter-annotator agreement provided; Realistic datasets (not toy); Other.

  \item \textbf{Writing \& Presentation}: 
  Clear problem definition; Well-structured and organized; Accessible language, minimal jargon; 
  Strong and informative figures \& tables; Consistent terminology; Grammatically correct, well-formatted; 
  Contribution clearly communicated; Other.

  \item \textbf{Broader Impact, Ethics \& Relevance}: 
  Thoughtful discussion of societal risks or ethical implications; Realistic and meaningful claims of impact; 
  Awareness of biases, fairness, or safety issues; Other.

  \item \textbf{References \& Related Work}: 
  Comprehensive related work; Covers important prior work; Up-to-date references; Other.

  \item \textbf{Fit \& Scope for Venue}: 
  Strong fit with venue; Other.

  \item \textbf{Other}: 
  Other.
\end{enumerate}

\medskip
\textbf{Output Format:}  
Return the annotations as a JSON list, where each element corresponds to a paragraph.  
Put the JSON list in the \texttt{<START\_LIST>} \texttt{<END\_LIST>}.

{\scriptsize
\begin{verbatim}
{
 "paragraph": "<original paragraph>",
 "annotation_list": [
   {
     "main_category": "<one of the 11 main categories>",
     "subcategory": "<one of the subcategories for the selected main category>",
     "source": "<exact phrase/sentence/paragraph that expresses the strength>"
   }
 ]
}
\end{verbatim}
}
\end{tcolorbox}
\caption{Strength annotation prompt}\label{fig:strength_prompt}
\end{figure*}

%% file: figures/argument_prompt.tex
\begin{figure*}[t]
\begin{tcolorbox}[sbox]
\textbf{Task Description:} You are an annotator.  

You are given a multi-turn conversation between reviewers and authors, starting with the reviewer.  
The reviewer message contains a set of weaknesses and questions.  
Your task is to analyze each reviewer weakness and question and the corresponding author response(s).  

For each reviewer weakness/question:  
\begin{enumerate}[leftmargin=1.5em, itemsep=0.2em]
  \item Determine if the author answered it, and record this in the \texttt{coverage} field.  
  If not answered, set \texttt{coverage} to \texttt{Not Answered} and leave all other fields empty.  

  \item Identify the author's stance towards the reviewer point in the \texttt{stance} field (\texttt{Disagree} or \texttt{Agree}).  

  \textbf{Agree/Disagree stance subcategories:}  
  \begin{itemize}[leftmargin=2em, itemsep=0pt]
    \item reject\_validity: Reject the validity of the question or weakness  
    \item evasion: Mitigate the importance of the question or weakness  
    \item reject\_request: Reject a request from the reviewer  
    \item contradict\_statement: Contradict a statement presented as a fact in the question or weakness 
    \item completion\_claim: Claim that a requested task has been completed  
    \item concede\_point: Concede the validity of a weakness or question  
    \item Promise a change by camera-ready deadline  
    \item future\_work: Express approval for a suggestion, but for future work  
  \end{itemize}

  \item Determine the author's strategy in answering (ignoring stance):  

  \textbf{evidence\_backed\_clarification subcategories:}  
  \begin{itemize}[leftmargin=2em, itemsep=0pt]
    \item method\_details: gives technical clarifications (formulas, hyperparameters, algorithm steps)  
    \item new\_table\_figures: provides new tables or figures  
    \item analysis: provides deeper breakdowns or error analysis  
    \item experiments\_in\_paper: refers to figures, tables, ablations, significance tests, baselines, or human evaluations in the original paper  
    \item new\_experiments: provides new ablations, tests, baselines, or evaluations done to answer the reviewer in rebuttal  
    \item citation: references prior work or other papers  
    \item other: evidence-backed but none of the above  
  \end{itemize}

  \textbf{generic\_defense\_vague subcategories:}  
  \begin{itemize}[leftmargin=2em, itemsep=0pt]
    \item repetition: repeats earlier claims without new support  
    \item broad\_assertion: general phrases like ``our method is strong'' without details  
    \item evasion: avoids addressing the reviewer’s concern directly  
    \item future\_promise: vague improvements promised without specifics  
    \item bare\_agreement\_or\_disagreement: only says ``we agree'' or ``we disagree'' without justification  
    \item other: vague or defensive but not fitting above  
  \end{itemize}
\end{enumerate}

\medskip
\textbf{Output Format:}  
Return the annotations as a JSON list inside \texttt{<START\_LIST>} \texttt{<END\_LIST>} tags.  
Each element corresponds to a reviewer weakness/question:  

{\scriptsize
\begin{verbatim}
{
 "reviewer_point": "<exact reviewer weakness/question>",
 "author_response": "<summary or quote of author response>",
 "coverage": "Answered / Not Answered",
 "stance": "<Disagree/Agree>",
 "stance_subcategory": "<stance subcategory>",
 "strategy": "<evidence_backed_clarification/generic_defense_vague>",
 "strategy_subcategory": "<strategy subcategory>"
}
\end{verbatim}
}
\end{tcolorbox}
\caption{Reviewer–Author dialogue annotation prompt}\label{fig:argument_prompt}
\end{figure*}